\documentclass[review]{elsarticle}

\usepackage{lineno}
\usepackage{hyperref}
\usepackage{graphicx}
\usepackage{textgreek}
\usepackage{amsmath}
\usepackage{microtype}
\usepackage{booktabs}
\usepackage{siunitx}
\usepackage{makecell}
\usepackage{footnote}
\usepackage{extdash}
\modulolinenumbers[2]

\journal{J. Quant. Spectrosc. Radiat. Transfer}

\begin{document}

\begin{frontmatter}

\title{The HITRAN2024 methane update}

\author[cfa]{T. Bertin\corref{cor1}}
\cortext[cor1]{Corresponding Author}
\ead{thibault.bertin@cfa.harvard.edu}

\author[cfa]{I. E. Gordon}

\author[cfa]{R. J. Hargreaves}

\author[ucl]{J. Tennyson}

\author[ucl]{S. N. Yurchenko}

\author[ucl]{K. Kefala}

\author[dijon]{V. Boudon}

\author[dijon]{C. Richard}

\author[iao]{A. V. Nikitin}

\author[tsu,iao]{V. G. Tyuterev}

\author[reims]{M. Rey}


\author[dlr]{M. Birk}

\author[dlr]{G. Wagner}

\author[jpl]{K. Sung}

\author[jpl,chic]{B. P. Coy}

\author[jpl,river]{W. Broussard}

\author[jpl]{G. C. Toon}

\author[iao]{A. A. Rodina}

\author[iao]{E. Starikova}

\author[gren]{A. Campargue}

\author[nist]{Z. D. Reed}

\author[nist]{J. T. Hodges}

\author[cfa,hefei]{Y. Tan}

\author[noaa,pld]{N. A. Malarich}

\author[pld]{G. B. Rieker}

\address[cfa]{Center for Astrophysics \textbar Harvard \& Smithsonian,  Atomic and Molecular Physics Division, Cambridge MA 02138, USA}
\address[ucl]{Department of Physics and Astronomy, University College London, London, WC1E 6BT, UK}
\address[dijon]{Universit{\'e} Bourgogne Europe, CNRS, Laboratoire Interdisciplinaire Carnot de Bourgogne ICB UMR 6303, F-21000 Dijon, France}
\address[iao]{V.E. Zuev Institute of Atmospheric Optics, Laboratory of Theoretical Spectroscopy, Russian Academy of Sciences, 634055 Tomsk, Russia}
\address[tsu]{National Research Tomsk State University, Physics Department, 634050 Tomsk, Russia}
\address[reims]{Groupe de Spectrom\'{e}trie Mol\'{e}culaire et Atmosph\'{e}rique, UMR CNRS 7331, BP 1039, F-51687, Reims Cedex 2, France}
\address[dlr]{German Aerospace Center (DLR), Remote Sensing Technology Institute, Wessling, Germany}
\address[jpl]{Jet Propulsion Laboratory, California Institute of Technology, Pasadena CA, USA}
\address[chic]{Department of the Geophysical Sciences, University of Chicago, Chicago IL, USA}
\address[river]{Department of Earth and Planetary Sciences, UC Riverside, Riverside CA, USA}
\address[gren]{University  of  Grenoble  Alpes,  CNRS,  LIPhy,  F-38000  Grenoble,  France}
\address[nist]{National Institute of Standards and Technology, Chemical Sciences Division, Gaithersburg MD, USA}
\address[hefei]{Hefei National Laboratory for Physical Science at Microscale, University of Science and Technology of China, Hefei, China}
\address[noaa]{National Oceanic and Atmospheric Administration, Chemical Sciences Laboratory, Boulder CO, USA}
\address[pld]{Precision Laser Diagnostics Laboratory, University of Colorado Boulder, Boulder CO, USA}

\begin{abstract}
Spectroscopic parameters of methane from many different studies were gathered to improve the HITRAN database towards its 2024 version. After a validation process using high-resolution FTS and CRDS spectra, about 80,000 lines of the four most abundant isotopologues were replaced from the dyad to the triacontad regions. These changes amount to 51,000 transition wavenumbers, 18,000 line intensities, 33,000 pressure-broadening half-widths, and 3300 assignments. 44,000 new lines were added with 16,000 old lines removed, extending the database from 12,000 cm$^{-1}$ up to 14,000 cm$^{-1}$, and covering some gaps. A greater focus was brought on the pentad, octad, and tetradecad regions, targeted by several remote sensing instruments. In these regions, comparisons of spectral fits from multiple line lists were performed, taking only the parameters that provide best fit for each line. In the $\nu _3$ band, in addition to replacing the previous values, speed-independent pressure broadening parameters of \textsuperscript{12}CH\textsubscript{4} were gathered and used to fit Padé-approximant functions. These functions then replaced any outdated experimental data in $\nu _3$, missing data in the new lines, as well as the values that were determined to be outside their physical boundaries. The CH\textsubscript{3}D broadening parameters were replaced in the same manner, for missing and low or high values, using a semi-empirical formula instead.
\end{abstract}

\end{frontmatter}

\section{Introduction}
\renewcommand{\color}[1]{}
The spectroscopy of methane has a fundamental role in the observations and radiative transfer modeling of gaseous media, especially atmospheres of Earth \cite{chanmiller_methane_2024}, solar system planets and moons \cite{10.1006/icar.1994.1139,10.1016/j.icarus.2005.02.004}, exoplanets (e.g. HD 102195b \cite{10.1051/0004-6361/201834615}), brown dwarfs (e.g. HD 102195b \cite{10.1093/MNRAS/STAC1412}), and combustion environments \cite{10.1364/AO.35.004026}. The recent development of remote observation tools such as those on board Sentinel-5 \cite{lorente_methane_2021}, GOSAT, GOSAT-2 \cite{imasu_greenhouse_2023}, and MethaneSAT \cite{chanmiller_methane_2024} has pushed the demand for higher quality parameters at many temperatures and pressures, in the presence of various other species, provided by the HITRAN database \cite{gordon_hitran2020_2022}. Methane is frequently measured in planetary and stellar atmospheres, where its study plays an important role in understanding the formation and composition of these astronomical objects. Important methane-specific data such as its presence, concentration, or partial pressure would not be retrievable without models that incorporate accurate spectroscopic parameters.


The octad (3500--5000 cm$^{-1}$) and tetradecad regions (5000--6250 cm$^{-1}$) in particular are crucial for satellite missions \cite{lorente_methane_2021,malina_consistency_2022,chesnokova_estimation_2020}, not only for its principal isotopologue but also for \textsuperscript{13}CH\textsubscript{4} and CH\textsubscript{3}D, which are used to determine \textsuperscript{13}C/\textsuperscript{12}C and D/H ratios. The octad has been improved in previous editions of HITRAN in the window observed by TROPOMI (4190--4340 cm$^{-1}$). However, at higher wavenumbers {\color{blue} of the octad}, spectral parameters were of lower accuracy, incomplete, or both. Current and planned CH\textsubscript{4} satellite sensors strive for ~0.5\% precision in the column (Ref. \cite{jacob_quantifying_2022}; Table 1) for area flux mapping purposes, which requires similar accuracies for the intensities and widths and is not achieved in many regions. Even in the TROPOMI window, the HITRAN uncertainty code only reaches 4 (i.e., $0.0001 \leq \Delta \nu < 0.001$ cm$^{-1}$ for line positions and $10\% \leq \frac{\Delta S_{ij}}{S_{ij}} < 20\%$ for intensities).

Above, and including the tetradecad region, other problems arise from the lack of assignment and/or a value for the lower state energy ($E''$). Given the temperature range, these issues are limited in the Earth's atmosphere, but make accurate spectral modeling impossible at more extreme temperatures (for example, low temperatures on Titan or high temperatures on hot-Jupiters and brown dwarfs). {\color{blue} With that one should note, that while for higher temperatures one can use the HITEMP version of the database \cite{10.3847/1538-4365/ab7a1a}, it also has limitations due to difficulty calculating accurate line positions at higher wavenumbers.} Furthermore, the origin of some of the parameters in HITRAN2020 is not well documented, especially for pressure-induced data {\color{blue} such as those being the results of private communications}. This can cause additional problems, preventing comparisons, or extrapolations and adjustments if the old data was generated by a model.

The focus of the HITRAN2024 update was on {\color{blue} the parameters in the 160-characters HITRAN format, while a few speed-dependent and first-order line-mixing parameters were added as well}. Significant progress has been made between 900 and 14000 cm$^{-1}$ \cite{kefala_empirical_2024,nikitin_improved_2024,rodina_improved_2021,nikitin_preliminary_2013,starikova_assignment_2024,malarich_dual_2021,nikitin_first_2024,campargue_high_2023} for the three most abundant isotopologues, \textsuperscript{12}CH\textsubscript{4}, \textsuperscript{13}CH\textsubscript{4}, and \textsuperscript{12}CH\textsubscript{3}D, allowing for an update to the transition wavenumber, intensity, and assignments from the HITRAN2020 database using various recorded spectra for validation. These line lists also cover missing portions of the infrared spectrum that were previously challenging to obtain, as well as some in-between lines that were missing. Similar efforts have been put into obtaining good quality pressure-induced line-shape parameters \cite{antony_n2_2008,bertin_co2_2024,clark_difference_2004,devi_self-_2015,devi_spectral_2016,devi_multispectrum_2018,es-sebbar_intensities_2014,es-sebbar_linestrengths_2021,farji_airinduced_2021,gabard_line_2010,gharib-nezhad_h2induced_2019,ghysels_spectroscopy_2011,ghysels_temperature_2014,grigoriev_estimation_2001,lyulin_measurements_2014,ma_temperature_2016,manne_determination_2017,mondelain_line_2005,mondelain_measurement_2007,pine_speeddependent_2019,pine_speeddependent_2000,pine_multispectrum_2003,richard_self_2023,dudaryonok_semiempirical_2018}, with the differences being that we considered older work, with the intention of replacing the parameters for more lines than those in Section \ref{sec:addition_shifts} to improve traceability. To this end, a combination of good quality measurements, with an empirical Padé-approximants model based on these measurements and semi-empirical models to fill the gaps was used. These results provide the necessary foundation for the inclusion of first-order and full line-mixing with speed-dependent profiles.

\section{Additions and updates {\color{blue}of non-pressure-induced data}}\label{sec:addition_shifts}
The newest versions of MARVELized-ExoMol \cite{yurchenko_exomol_2024}, TheoReTS \citep{10.3847/1538-4357/aa8909}, HITEMP  \citep{10.3847/1538-4365/ab7a1a} and MeCaSDa \cite{MeCaSDa_2013}, which is partly accessible via VAMDC portal \cite{rixon_VAMDC_2011, albert_VAMDC_2020}, line lists as well as lists from Nikitin \textit{et al.} \cite{nikitin_improved_2024,nikitin_preliminary_2013,nikitin_first_2024}, Rodina \textit{et al.} \cite{rodina_improved_2021}, Sung \textit{et al.} \cite{sung_new_2024}, Starikova \textit{et al.} \cite{starikova_assignment_2024}, Malarich \textit{et al.} \cite{malarich_dual_2021}, and Campargue \textit{et al.} \cite{campargue_high_2023} were available for the HITRAN2024 update and are summarized for each region covered. These works were then validated against several high-quality, high-resolution FTS \cite{2015ApJ...813...12H,bertin_co2_2024,birk_esa_2017,birk_measurement_2025} and CRDS spectra \cite{birk_esa_2017} obtained under well-controlled laboratory conditions, when available for the updated region. The best parameters depending on the spectral region were selected and added to the HITRAN2024 database.

Much of the focus was brought to the pentad, octad, and tetradecad regions, but many other spectral regions have also been updated for HITRAN2024. Changes from these lists were applied almost exclusively for the transition wavenumbers, intensities, and energy levels, since pressure-induced parameters of most of the available line lists were missing or approximate. Updates on these parameters are detailed in Sec. \ref{sec:pressure_induced}. The newly added lines, which were formerly missing in HITRAN, were added as is, but their pressure broadenings and shifts were replaced as well.

    {\color{blue}\subsection{Full spectrum lists at disposal}\label{sec:provided_linelists}
    Other compiled lists of calculated or measured line parameters were available for this update and summarized here. These were either included in other work or directly taken in some regions of HITRAN2024 as described starting Sec. \ref{sec:900-1050}.} 

        \subsubsection{ExoMol}
        The ExoMol project \cite{tennyson_2024_2024} has produced a number of methane line lists of increasing completeness and accuracy designed for use at high temperatures \cite{yurchenko_exomol_2014,yurchenko_hybrid_2017,yurchenko_exomol_2024}. The intensities in this list are calculated variationally. The line positions in the most recent of these \cite{yurchenko_exomol_2024}, known as the MM line list, was based on a comprehensive MARVEL (measured active vibration-rotation energy levels) study of \textsuperscript{12}CH\textsubscript{4} by Kefala \textit{et al} \cite{kefala_empirical_2024}. This MARVEL study extracted  82173 rovibrational and rotational transitions from 96 literature sources containing high-resolution studies of \textsuperscript{12}CH\textsubscript{4} spectra; the study identified a further 42 sources containing unassigned or partially assigned methane spectra which could not be included in the MARVEL study. An analysis of these sources is currently in progress and has so far led to the assignment of almost 10000 further lines, which extend the coverage of the previous study in terms of coverage of both rotational and vibrational states. This work can enable updates to future releases of both HITRAN and HITEMP. A HITRAN-format 296~K version of the MM line list generated using PyExoCross \cite{Zhang_pyExoCross_2024} was used in this study, {\color{blue}predominantly in the pentad region.}

        \subsubsection{TheoReTS}
        The TheoReTS database \cite{rey_theorets_2016}  provides line lists and quasi-continuum theoretical calculations for low and high temperatures from \textit{ab initio} potential energy surfaces (PES) and dipole moment surfaces (DMS) together with spectra simulation tools. It was recently updated in  \citep{10.3847/1538-4357/aa8909} \cite{rey_new_2018}.
        Variational line position calculations by Rey \textit{et al.} have been refined using available empirical energy levels deduced from effective spectroscopic models. The TheoReTS predictions have been experimentally validated via high-temperature laser experiments by Ghysels \textit{et al.} \cite{ghysels_temp_2018} and by Wong \textit{et al.} \cite{Wong_atlas_2019} for their methane cross sections,  as well as in dual frequency comb experiments  and double resonance experiments by Malarich \textit{et al.} \cite{malarich_dual_2021} and Foltynowicz \textit{et al.} \cite{foltynowicz_2021} respectively. They were also used by Rey \textit{et al.} \cite{rey_new_2018} to model methane absorption at Titan, as observed by VIMS and by DISR instruments onboard the Cassini and Huygens space missions, and for the interpretation of non-LTE experiments in a hypersonic jet flow by Dudas \textit{et al.} \cite{dudas_2023}.
        In the context of the present HITRAN2024 update,  TheoReTS predictions were used to create the HITEMP methane line list \citep{10.3847/1538-4365/ab7a1a}, while the extended assignments of \textsuperscript{12}CH\textsubscript{4}, \textsuperscript{13}CH\textsubscript{4} and \textsuperscript{12}CH\textsubscript{3}D experimental spectra were carried out with improved line lists as described in the following sections.  In recent multi-laboratory line intensity measurements, Reed \textit{et al.} \cite{reed_multilaboratory_2025} reported  permille-level uncertainties of the R(0) to R(10) manifolds in the CH\textsubscript{4} $2\nu_3$ band. These results had an average agreement of about 0.05 \% with TheoReTS calculations which used the \textit{ab initio} DMS of ref. \cite{NIKITIN_dipole_intens_2017}.

        \subsubsection{HITEMP}
        The HITEMP spectroscopic database \citep{10.1016/j.jqsrt.2010.05.001} contains line lists for eight molecules, with the methane line list fully described in Hargreaves \textit{et al.} \citep{10.3847/1538-4365/ab7a1a}. The HITEMP list was constructed by combining the empirically corrected \textit{ab initio} line list for \textsuperscript{12}CH\textsubscript{4} from TheoReTS \citep{10.3847/1538-4357/aa8909} with the HITRAN2016 line list \citep{gordon_hitran2016_2017}, which was the contemporary HITRAN list at the time. HITEMP CH$_{4}$  spans 0-13,400~cm$^{-1}$ and is suitable for simulating the high-temperature spectrum of methane up to 2000~K. In addition, the HITEMP list includes ``effective'' lines that are able to reproduce the continuum-like features provided by temperature-dependent lists of weak lines (i.e.,  so-called super-lines) in TheoReTS. Readers are referred to Hargreaves \textit{et al.} \citep{10.3847/1538-4365/ab7a1a} for a full description of the ``effective'' line methodology for methane. 
        
        Merging HITRAN2016 with the TheoReTS line list required consistency with assignments, as well as positions and intensities to ensure uniqueness of transitions. Below 6230 ~cm$^{-1}$, the HITRAN2016 line list contained both upper-state and lower-state assignments for vibration and rotation quanta; however, above this point, the assignments were limited. The assignments of the TheoReTS list were also incomplete, which enabled lines to be matched only by their positions, intensities, lower state energies, and rotation number ($J''$). Consequently, approximately 81,000 ($\sim$50\%) of $^{12}$CH$_{4}$ lines in HITRAN2016 replaced corresponding lines of TheoReTS. Furthermore, there were no equivalent high temperature line lists for isotopologues of methane ($^{13}$CH$_{4}$, $^{12}$CH$_{3}$D, $^{13}$CH$_{3}$D) and these were imported into HITEMP directly from HITRAN2016. The resultant HITEMP line list contains $\sim$32 million lines, of which approximately 5 million are effective lines and for which the largest contribution of lines originates from the $^{12}$CH$_{4}$ line list of TheoReTS. 

        \subsubsection{MeCaSDa}
        CaSDa (Calculated Spectroscopic Databases) \cite{RICHARD2024109127} is a project that presently gathers ten databases of line lists for ten highly symmetrical molecules including methane. These lists are built upon the analysis of high-resolution spectra (microwave, infrared, and Raman) using effective Hamiltonian and dipole moment operators written with a tensorial formalism \cite{BOUDON2004620}. The databases also contain the full description of eigenvectors.    
        Within this project, the MeCaSDa database was built for ${}^{12}$CH$_4$ and ${}^{13}$CH$_4$. The last update of MeCaSDa was performed in 2020 \cite{RICHARD2020107096}, following the results of a global fit described in Ref. \cite{10.1063/1.5023331}.

    {\color{blue}\subsection{900--1050 \texorpdfstring{cm$^{-1}$}{cm-1}}\label{sec:900-1050}}
        \subsubsection{IAO FTIR measurements}
        Four spectra of methane samples at natural isotopic abundance were recorded between 900--1050 cm$^{-1}$ at the Institute of Atmospheric Optics (IAO) in Tomsk \cite{nikitin_improved_2024}. Nonlinear least-squares curve fitting procedures were used to retrieve line positions and intensities, which were subsequently analyzed using effective Hamiltonian and effective dipole moment models. The enhanced absorption at long paths allowed the measurement of new transitions: up to $J = 28$ for cold bands and up to $J = 20$ for hot bands. The new experimental list contains line positions and intensities for 2570 lines, with quantum assignments made for 1244 of these lines for the main isotopologue \textsuperscript{12}CH\textsubscript{4}. Comparisons of theoretical absorption simulations with experimental spectra revealed a considerable improvement compared to the HITRAN2020 database. All of the 1244 assigned line positions were fit with an average RMS standard deviation equal to 0.00065 cm$^{-1}$.

        {\color{blue}\subsubsection{HITRAN update in the 900--1050 \texorpdfstring{cm$^{-1}$}{cm-1} region}\label{sec:hit_900-1050}}
        The new line list from Nikitin \textit{et al.} \cite{nikitin_improved_2024} (\textsuperscript{12}CH\textsubscript{4}, \textsuperscript{13}CH\textsubscript{4},  \textsuperscript{12}CH\textsubscript{3}D \& \textsuperscript{13}CH\textsubscript{3}D) completes the lower energy part of the dyad region, adding many lines including those for the lower-abundance \textsuperscript{13}CH\textsubscript{3}D isotopologue. Their data were merged directly to the region, replacing the transition wavenumbers for any difference ($\Delta \tilde{\nu} = 0$ cm$^{-1}$) between HITRAN2020 and the new parameters, the lower energy levels for $\Delta E''= 0.1$ cm$^{-1}$, and the intensities for $\Delta S_{ij} / S_{ij} = 0.1$. About half of the lines in the list mainly served to update the positions and intensities with a few energy levels and assignments, while the other half was entirely new.
        
    {\color{blue}\subsection{2500--3900 \texorpdfstring{cm$^{-1}$}{cm-1}}\label{sec:2500-3900}}
        {\color{blue}\subsubsection{HITRAN update in the 2500--3900 \texorpdfstring{cm$^{-1}$}{cm-1} region}}
        The pentad region is relatively well defined in HITRAN2020, nevertheless, we have incorporated a few noticeable improvements. Spectra from Hargreaves \textit{et al.} \citep{2015ApJ...813...12H} ($S1$) and Bertin \& Vander Auwera \cite{bertin_co2_2024} ($S2$) were used to validate the spectral parameters. Although $S1$ spans a much larger region (2500--4900 cm$^{-1}$), $S2$ (2860--3160 cm$^{-1}$) provides a higher signal-to-noise ratio (SNR) and higher resolution, making it a better choice for validating the line list in this window. Small missing parts appear as well in $S1$ from cutting water absorption; making these small windows unusable for validation.
    
        \begin{figure}[!htbp]
            \includegraphics[width = \textwidth]{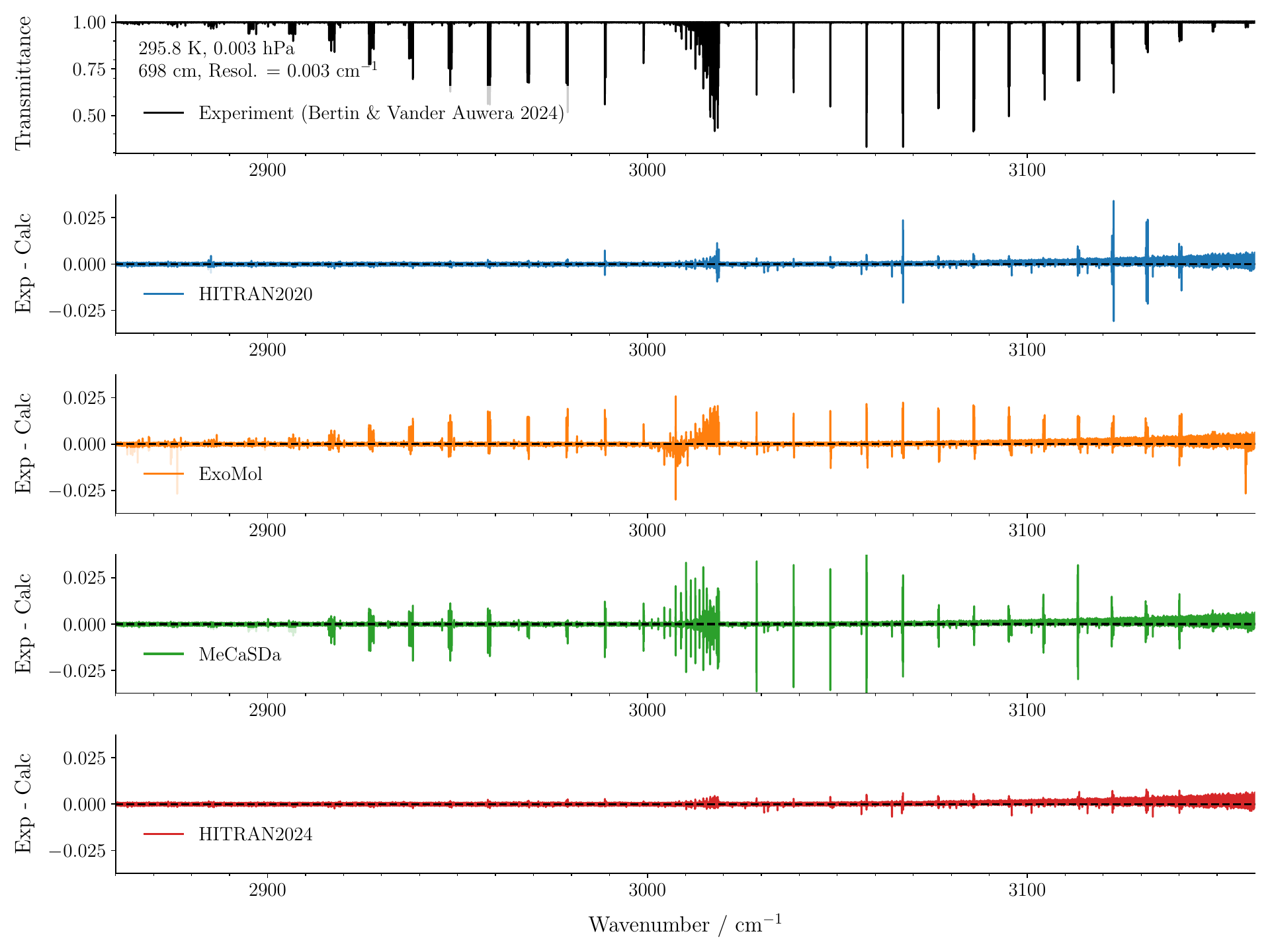}
            \caption{Residuals obtained at 296 K between experiment from $S2$ Bertin \& Vander Auwera \cite{bertin_co2_2024} and Voigt calculations from the line lists used for the HITRAN2024 update}
            \label{fig:bertin_sources}
        \end{figure}
    
        The residuals between $S1$, $S2$ and the models calculated using the HITRAN Application Programming Interface (HAPI) with lines from ExoMol (\textsuperscript{12}CH\textsubscript{4}), MeCaSDa (\textsuperscript{12}CH\textsubscript{4} \& \textsuperscript{13}CH\textsubscript{4}), and HITEMP (\textsuperscript{12}CH\textsubscript{4} \& \textsuperscript{13}CH\textsubscript{4}) were used to compare the quality of each parameter on a line-by-line basis (these residuals are presended for $S2$ in Fig. \ref{fig:bertin_sources}). HITEMP was used when all other data were insufficient, since many lines are not fully assigned. By artificially increasing the pressure of the models, we could isolate the missing or strongly shifted lines in every line list, resulting in downward-pointing peaks. Using a combination of derivatives and thresholding, we could automatically find which areas were missing in HITRAN and which line lists could cover them. Because the residuals can be caused by multiple missing lines, every line above a certain intensity in a small window around the region to update was taken from the new line list. We then used the same technique, without the pressure increase, to find and update the weakly shifted lines in HITRAN. In addition to this automated approach, manual selections were made between lists for lines that were incorrectly updated or missed because of the various thresholds.

        {\color{blue} Using the same HITRAN2020 residuals, some lines were found to have a worse quality of fit than their counterparts back from HITRAN2000. These transition wavenumbers and intensities were manually reverted to their 2000 values.}

    {\color{blue}\subsection{3900--4900 \texorpdfstring{cm$^{-1}$}{cm-1}}\label{sec:3900-4900}}
        \subsubsection{Kitt Peak \& JPL FTS measurements of \texorpdfstring{\textsuperscript{13}CH\textsubscript{4}}{13CH4}}
        Multiple sets of \textsuperscript{13}CH\textsubscript{4} line lists, either empirical or model predictions, were available from various sources, e.g., HITRAN 2020 \cite{gordon_hitran2020_2022,brown_measurements_2016}, JPL GFIT package (“atm.161”) \cite{toon_spectrometric_2021}, and Kitt Peak FTS (291~K) and JPl-FTS (80~K). More importantly, comprehensive updates to the previous work by Brown \textit{et al.} \cite{brown_measurements_2016} became available through the effective Hamiltonian modeling extended to the tetradecad region for hot band transitions. All of the line lists have been evaluated and cross-compared against high-resolution laboratory spectra obtained at 291 K and 80 K {\color{blue} by generating synthetic spectra for the the same experimental conditions (e.g. sample pressure, temperature, gas cell pathlength, and spectral resolution) as the corresponding laboratory spectra, but using different sets of line lists available. These multiple sets of synthetic spectra were compared with lab spectra at each temperature to determine} their quantum identifications and lower state energy values {\color{blue} by adopting model calculation} as presented in Sung \textit{et al.} \cite{sung_new_2024}
    
        More specifically, by adopting a two-temperature method, the updates were first carried out by generating synthetic spectra at 296 K and 80 K based on each of the aforementioned line lists under the same experimental conditions with the corresponding laboratory spectra (one from Kitt Peak FTS at 291 K and the other from JPL-FTS at 80 K). Methodical line-by-line comparisons were done for all individual features observed in the laboratory spectra. In doing so, any transitions of residual H\textsubscript{2}O and \textsuperscript{12}CH\textsubscript{4} were easily excluded as much as they were identifiable in reference to the HITRAN database. Once their corresponding line pairs between synthetic and observed features were identified by visual inspection assisted with user-interactive graphic tools {\color{blue}(designed and developed in Matlab for line matching purpose)}, a new line list was generated by piecing together the best values per line parameters. For any observed features having no acceptable match in either the model predictions or in the HITRAN2020 database \cite{gordon_hitran2020_2022,brown_measurements_2016} within a reasonable offset from their line positions and intensities, their lower state energies were intentionally assigned an estimated value (e.g. 444.4444, 120.000, etc) {\color{blue} only implying that the transition belongs to low J-transitions or not}. For the sake of completeness, we also assumed air- and self-widths to be 0.055 and 0.065 cm$^{-1}/$atm at 296 K with the temperature-dependent exponent to be 0.7 for the air-widths unless their measurement values were available in HITRAN2020 \cite{gordon_hitran2020_2022} or atm.161 \cite{toon_spectrometric_2021}. 
        
        \begin{table}[!htbp]
            \centering
            \caption{Statistics on the line parameters in comparison to the existing spectroscopy databases and model calculations for \textsuperscript{13}CH\textsubscript{4}.}
            \label{tab:sung_13ch4stat}
            \begin{tabular}{lllp{25ex}}
                \toprule
                \textsuperscript{13}CH\textsubscript{4} line list & Range & Min. ($S_{ij}$) &lines listed/ assigned\\
                \midrule
                HITRAN2020 & 4000--4700 cm$^{-1}$ & $1 \times 10^{-26}$ cm$^{-1}$atm$^{-1}$ & 761 / 693\\
                Coy, Bourssard, Toon, Sung \cite{sung_new_2024} & 4000--4700 cm$^{-1}$ & $1 \times 10^{-26}$ cm$^{-1}$atm$^{-1}$ & 7103 /  6585\\
                \bottomrule
            \end{tabular}
        \end{table}
        
        Table \ref{tab:sung_13ch4stat} presents statistics on the improvements (both additions and updates inclusive) as reported by Sung \textit{et al.} \cite{sung_new_2024}, showing an almost ten-fold increase in the total number of line list entries in the 4000--4700 cm$^{-1}$ wavenumber interval for \textsuperscript{13}CH\textsubscript{4}, when compared to the HITRAN2020.
        \begin{figure}[!htbp]
            \centering
            \includegraphics[width=1.0\linewidth]{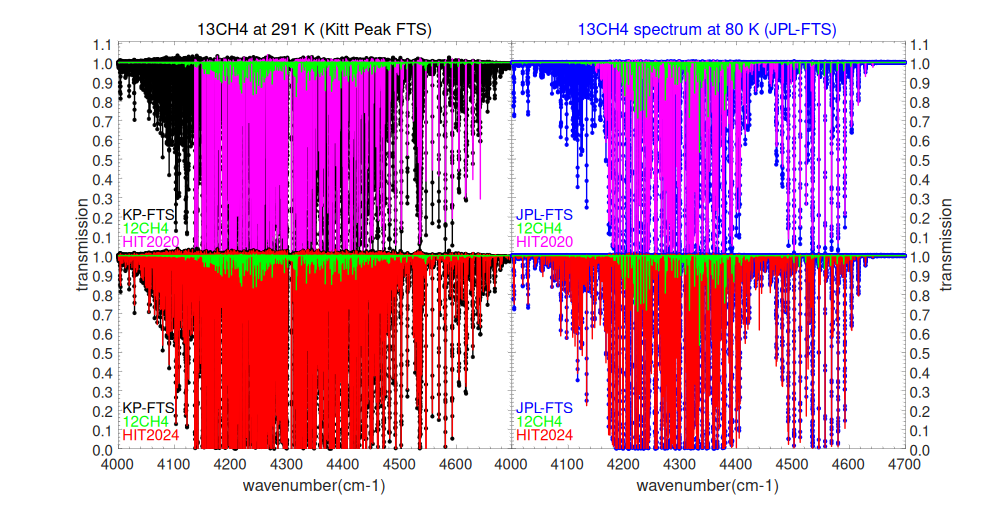}
            \caption{Comparison of synthetic spectra generated with HITRAN2020 (in pink) and our updated composite line list (in red) with high-resolution Kitt Peak FTS spectrum at 291 K (Left, in black) and JPL-FTS spectrum at 80 K (Right, in blue). This plots show better performance and broader spectral coverage by the new composite line list. Note that the synthetic spectra were computed for the same conditions as the corresponding laboratory spectra. Any H\textsubscript{2}O (not presented) and \textsuperscript{12}CH\textsubscript{4} features (in green) were identified during the line-by-line matching, as presented for \textsuperscript{12}CH\textsubscript{4} features.}
            \label{fig:sung_13ch4octad}
        \end{figure}
        Finally, Fig. \ref{fig:sung_13ch4octad} presents a graphical summary of the updates, showing synthetic spectra generated with the HITRAN2020 line list and the composite line list reported in this work compared to two high-resolution laboratory FTIR spectra, respectively at 291 K( Left panel) and 80 K (Right panel).  The synthetic spectra at 291 K and 80 K were computed for the same conditions as the corresponding laboratory spectra. It is shown that the composite line list covers much broader section of the Octad region as well as providing a better representation of the observed spectra at both temperatures thanks to the lower-state energy updates.

        {\color{blue}\subsubsection{HITRAN update in the 3900--4900 \texorpdfstring{cm$^{-1}$}{cm-1} region}}
        Spectra $S1$ and from Birk \textit{et al.} \cite{birk_esa_2017} ($S3$), including CRDS spectra not shown here, were used for the octad region. For the same reason {\color{blue} that $S1$ shows a lower SNR}, $S3$ was used for its window of 3900--4500 cm$^{-1}$ instead.
            
        \begin{figure}[!htbp]
            \includegraphics[width = \textwidth]{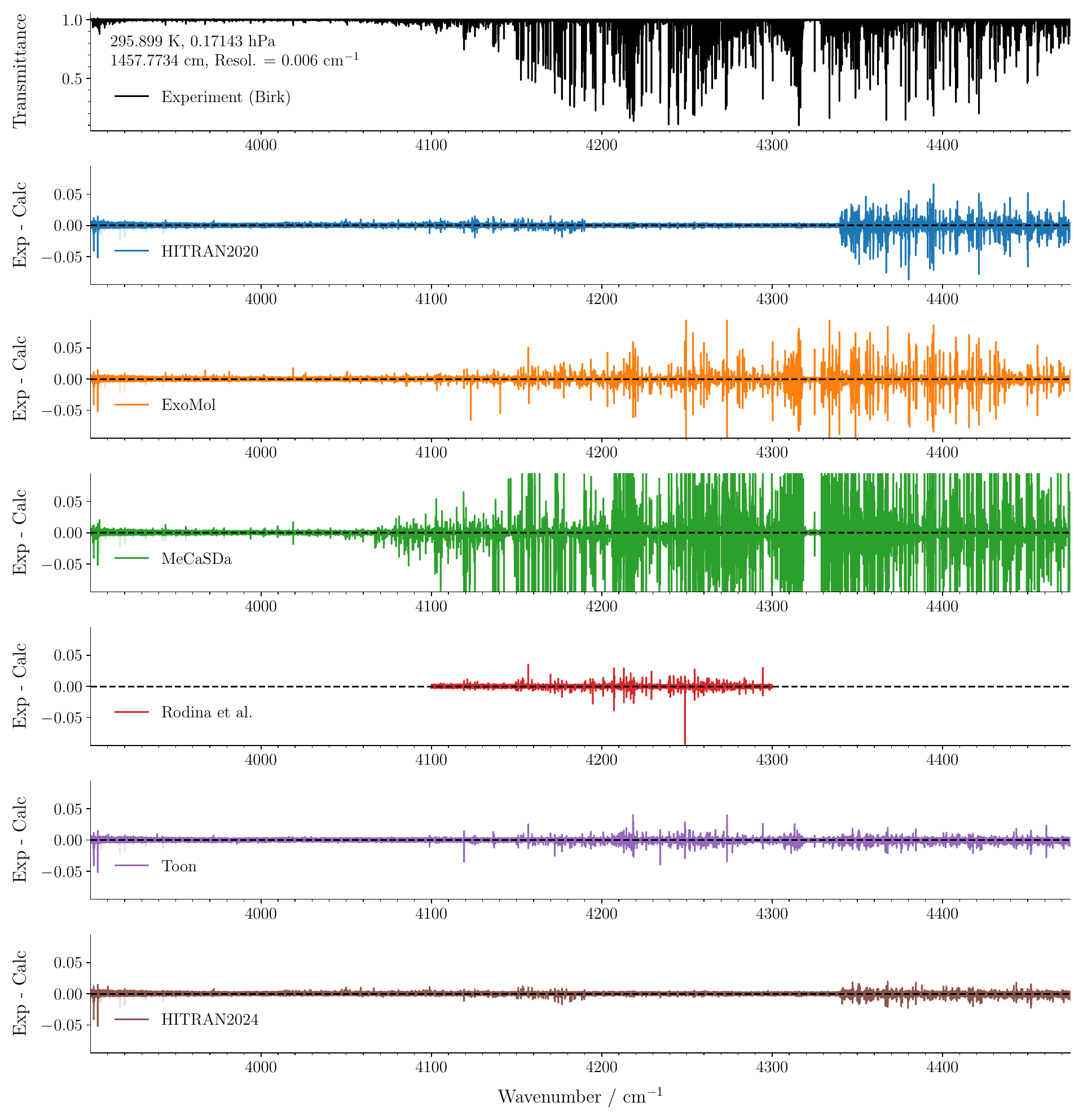}
            \caption{Residuals obtained at 296 K and 0.17143 hPa between the experiment $S3$ from Ref. \cite{birk_esa_2017} and Voigt calculations from the line lists used for the HITRAN2024 update.}
            \label{fig:birk_sources}
        \end{figure}
    
        \begin{figure}[!htbp]
            \includegraphics[width = \textwidth]{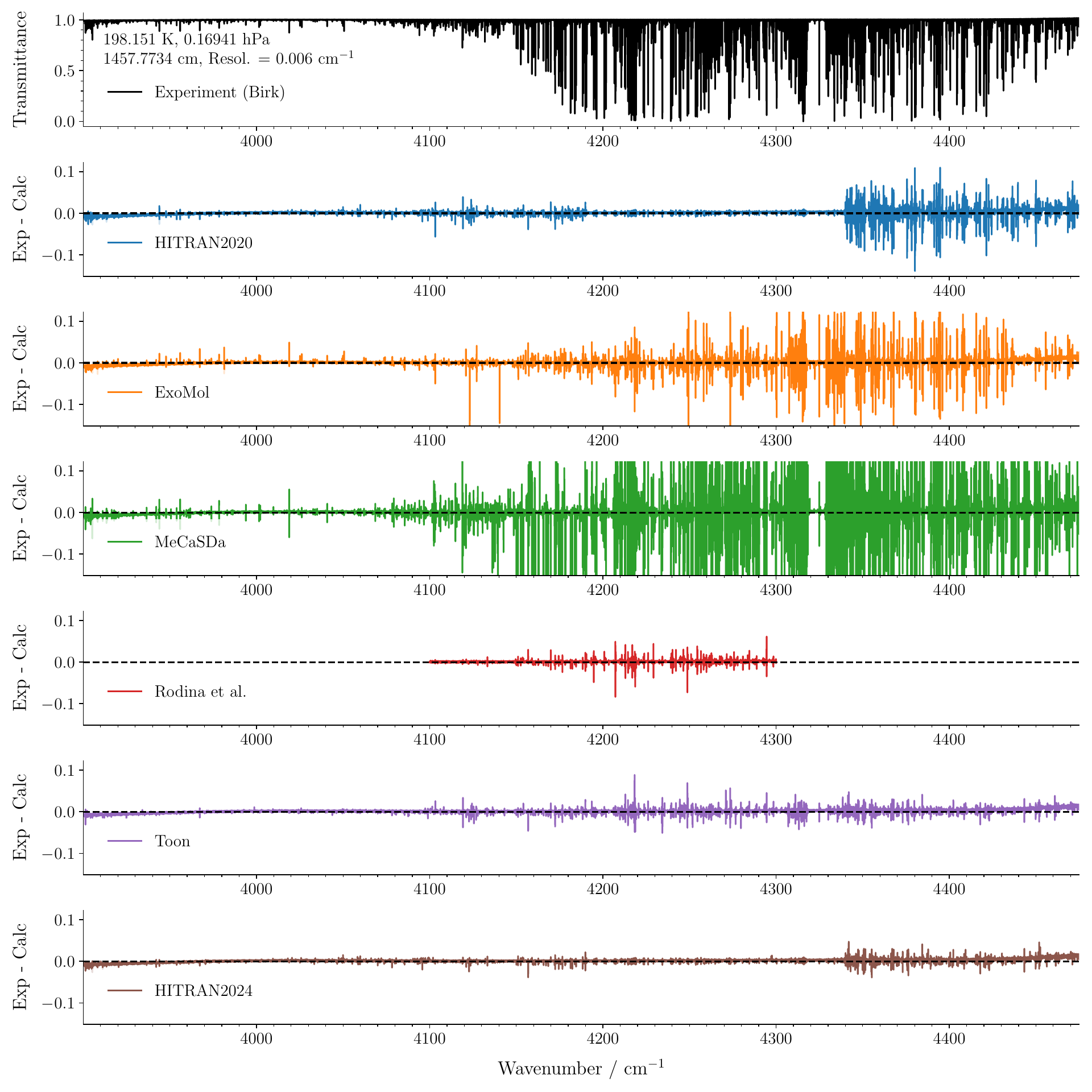}
            \caption{Residuals obtained at 198 K and 0.16941 hPa between the experiment $S3$ from Ref. \cite{birk_esa_2017} and Voigt calculations from the line lists used for the HITRAN2024 update.}
            \label{fig:birk_sources_198k}
        \end{figure}
    
        \begin{figure}[!htbp]
            \includegraphics[width = \textwidth]{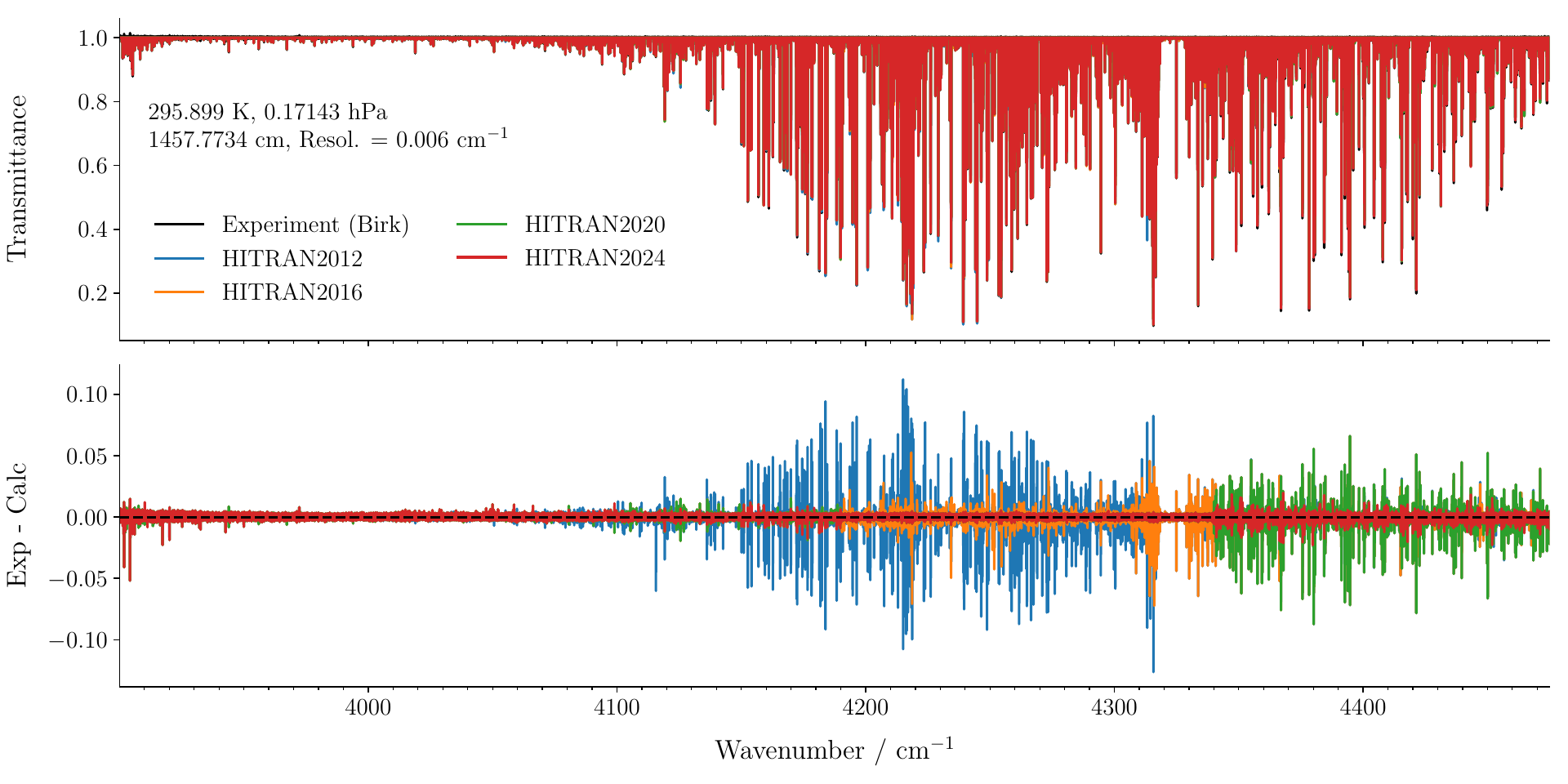}
            \caption{Residuals obtained between experiment $S3$ from Birk and Voigt calculations from the past and new HITRAN versions}
            \label{fig:birk_hitran}
        \end{figure}
        
        Unlike in the pentad region, several new line lists presented a clear improvement over large portions of the spectrum, as illustrated in Figures \ref{fig:birk_sources} and \ref{fig:birk_sources_198k}. This result alleviated the need for line-by-line analysis, with a few exceptions that were then fixed. In these {\color{blue} large portions}, the new lists {\color{blue} were merged with} HITRAN2020, adding new lines and updating existing ones unless no significant changes were made by conditions {\color{blue} similar to those} detailed in Sec. \ref{sec:hit_900-1050} {,\color{blue} depending on the precision of the parameters}. The residuals were then analyzed, keeping HITRAN2020 parameters if needed, ensuring that only the best parameters remained. This method was used for the data from Rodina \textit{et al.} \cite{rodina_improved_2021}, between 4100 and 4190 cm$^{-1}$, improving both CH\textsubscript{4}, \textsuperscript{13}CH\textsubscript{4}, and CH\textsubscript{3}D. Their line list extended to 4300 cm$^{-1}$, but that region was previously updated in HITRAN2020 for TROPOMI \cite{hu_operational_2016}. A list provided by Sung \textit{et al.} for \textsuperscript{13}CH\textsubscript{4} has been added as well between 4000--4700 cm$^{-1}$, including the region between 4190--4300 cm$^{-1}$ which replaces some lines, or intensities, incorrectly assigned to the main isotopologues. The list from Nikitin \textit{et al.} \cite{nikitin_preliminary_2013} improving CH\textsubscript{3}D between 4000--4554 cm$^{-1}$ has been added, adding the lines missing in HITRAN2016 \cite{gordon_hitran2016_2017} when the preliminary version of their data was first used. Finally, the empirical list from Toon (atm.161 {\cite{toon_atmospheric_2022}) {\color{blue} replaced the region} between 4341--4900 cm$^{-1}$ {\color{blue} for lines above the intensity of $10^{-27}$ cm$/$molecule}, improving CH\textsubscript{4}, \textsuperscript{13}CH\textsubscript{4}, and CH\textsubscript{3}D.
    
        We observed an approximately systematic shift in the transition wavenumber of lines from Daumont \textit{et al.} \cite{daumont_new_2013}. Applying a constant shift of 0.001 cm$^{-1}$ appeared to improve the residuals, particularly below 4200 cm$^{-1}$. The effect is limited above 4375 cm$^{-1}$ as the addition of atm.161 list already improved that region appreciably. The final residuals for the latest HITRAN versions are shown in Figure \ref{fig:birk_hitran}, confirming the great improvement above 4375 cm$^{-1}$, after keeping the region unchanged in the last 3 versions.

    {\color{blue}\subsection{4970--5300 \texorpdfstring{cm$^{-1}$}{cm-1}}\label{sec:4970-5300}}
        \subsubsection{IAO analyses of JPL FTS measurements of \texorpdfstring{\textsuperscript{13}CH\textsubscript{4}}{13CH4}}
        \begin{figure}[!htbp]
            \centering
            \includegraphics[width=0.75\linewidth]{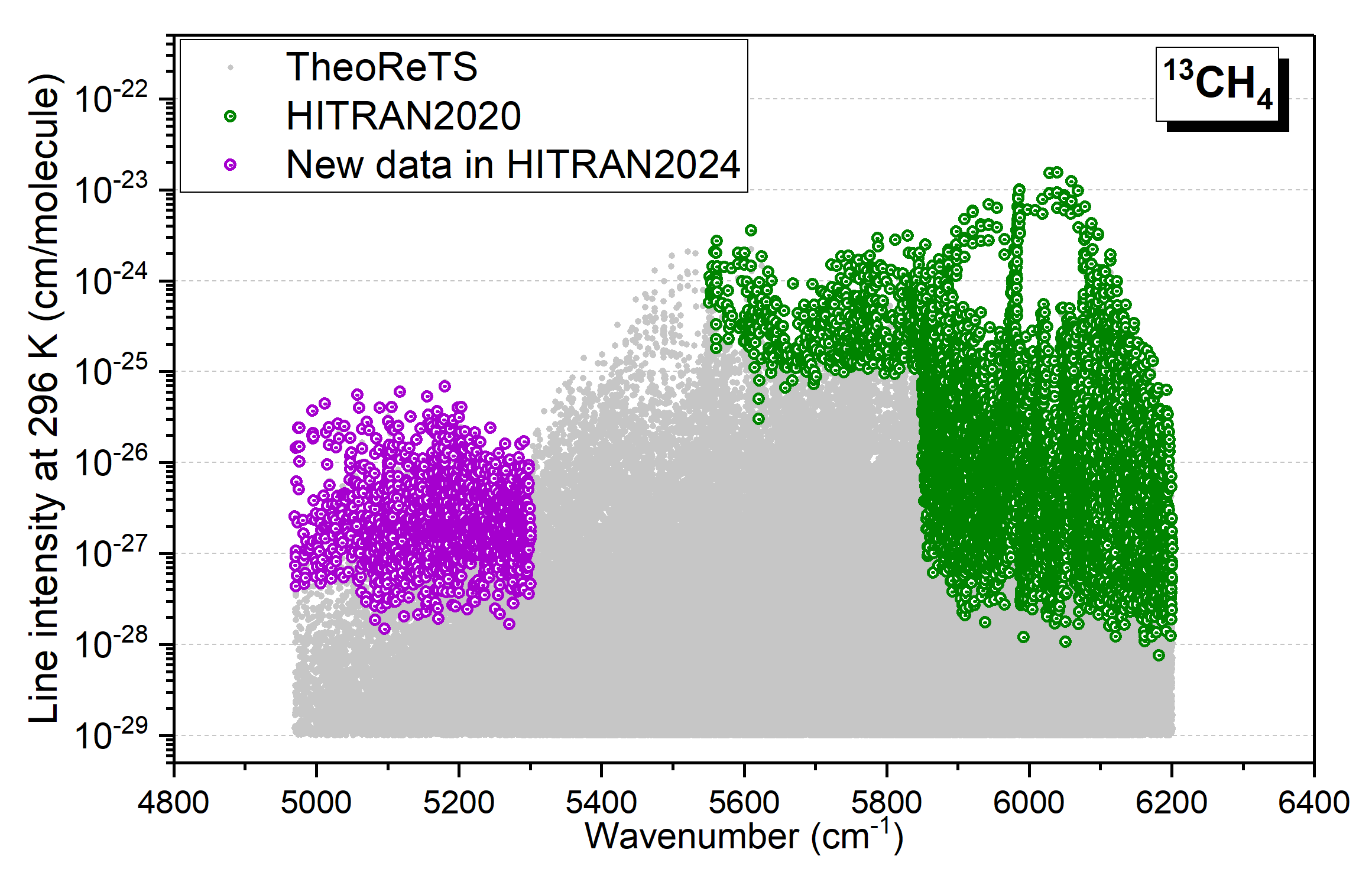}
            \caption{An overview of the \textsuperscript{13}CH\textsubscript{4} tetradecad line lists available in HITRAN2024. Newly added line parameters for \textsuperscript{13}CH\textsubscript{4} are highlighted in magenta. The TheoReTS \textit{ab initio} based variational list \cite{rey_new_2018} with intensity cut-off $1 \times 10^{-29}$ cm/molecule (grey dots) is plotted as a background overlapping with the colored data}
            \label{fig:new_13ch4_tetradecad}
        \end{figure}
        Starikova \textit{et al.} \cite{starikova_assignment_2024} reported the first assignments of \textsuperscript{13}CH\textsubscript{4} absorption bands belonging to the lower part of the tetradecad. This work focused on the analysis of a high-resolution Fourier-transform infrared spectrum of \textsuperscript{13}C-enriched methane in the 4970--5300 cm$^{-1}$ range (Fig. \ref{fig:new_13ch4_tetradecad}). 
        The spectrum was recorded over a broad range (4970--6200 cm$^{-1}$) at 298 K using a 20.94 m path cryogenic Herriott cell vacuum-coupled to a Fourier transform spectrometer, Bruker IFS-125HR, at the Jet Propulsion Laboratory. The retrievals of line intensities from the experimental spectrum were carried out using the SpectraPlot \cite{nikitin_visualization_2011} program, with the Voigt line profile at fixed values  $\gamma _\mathrm{self}= 0.08\  \mathrm{cm}^{-1}\mathrm{atm}^{-1}$, $n_\mathrm{self} = 0.85$, and $\delta _\mathrm{self} = - 0.018\ \mathrm{cm}^{-1} \mathrm{atm}^{-1}$. The uncertainties in the intensities of isolated lines were estimated to be about 10--15\%, while those of blended lines were about 20--40\%.
        Initial spectral assignments were obtained using the effective Hamiltonian model and effective dipole transition moment parameters, both derived from the previous work on cold \textsuperscript{13}CH\textsubscript{4} spectra recorded at 80 K \cite{starikova_13ch4_2018,starikova_assignment_2019,starikova_assignment_2016}.  The published experimental list contains 1642 lines, of which 1600 transitions were assigned to four vibrational bands $4\nu _4$, $\nu _2 + 3\nu _4$, $\nu _3 + 2\nu _4$ and $\nu _1 + 2\nu _4$ of the tetradecad system, including 22 sub-bands with rotational quantum numbers up to $J_\mathrm{max} = 16$. The parameters of the effective Hamiltonian and the dipole transition moment allowed the description of 1548 fitted experimental line positions with an rms deviation of $1.7 \times 10^{-3} \mathrm{cm}^{-1}$, and the reproduction of a set of 674 measured line intensities with an rms deviation of 6.8\%.
        \begin{figure}[!htbp]
            \centering
            \includegraphics[width=0.80\linewidth]{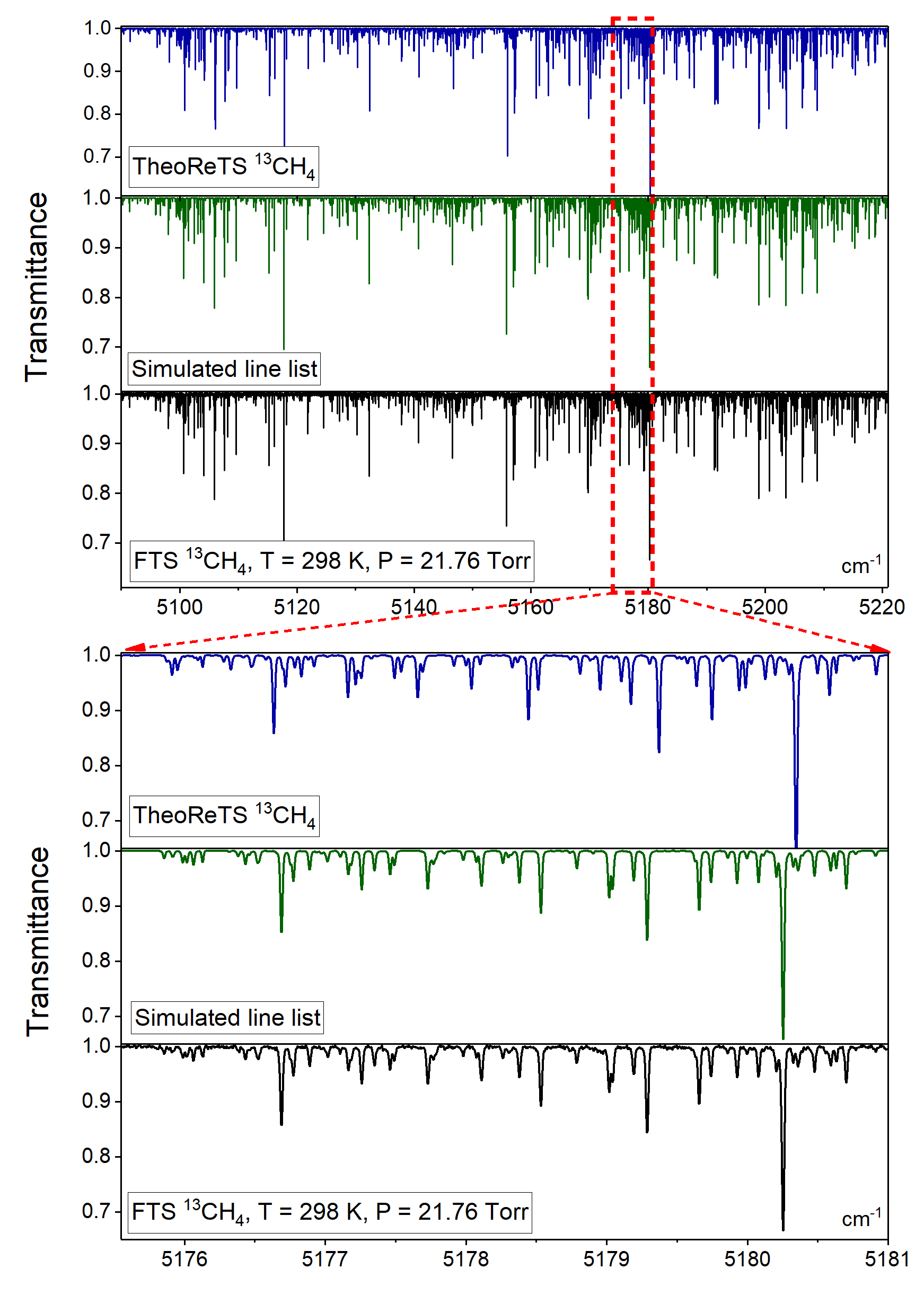}
            \caption{Comparison between the experimental spectrum of \textsuperscript{13}CH\textsubscript{4} at 298 K (lower panel) with simulation based on the effective Hamiltonian and dipole transition moment models (middle panel) and \textit{ab initio} TheoReTS \cite{rey_new_2018} calculations (upper panel) in the 5000--5220 cm$^{-1}$ tetradecad region. The blown-up part at the bottom corresponds to the narrow spectral interval around 5178 cm$^{-1}$.}
            \label{fig:spectra_13ch4_tetradecad}
        \end{figure}
        A comparison of the observed spectrum with the simulations using calculated effective parameters and the TheoReTS \cite{rey_new_2018} line list, depicted in Fig. \ref{fig:spectra_13ch4_tetradecad}, shows the excellent quality of the calculations for this molecule as well as the improvements due to the reduction of experimental data.

        {\color{blue}\subsubsection{HITRAN update in the 4970--5300 \texorpdfstring{cm$^{-1}$}{cm-1} region}}
        The new data from Starikova \textit{et al.} \cite {starikova_assignment_2024} partially fills the gap of \textsuperscript{13}CH\textsubscript{4} in the lower part of the tetradecad region. More than 1600 new lines were incorporated in HITRAN2024 (see Fig. \ref{fig:new_13ch4_tetradecad}). No \textsuperscript{13}CH\textsubscript{4} lines were present in this region in HITRAN2020.

    {\color{blue}\subsection{5900--6530 \texorpdfstring{cm$^{-1}$}{cm-1}}\label{sec:5900-6530}}
        \subsubsection{NIST, DLR \& USTC measurements}
        The tetradecad spectral region has been the subject of extensive study, including the WKLMC empirical line list \cite{campargue_2013}, the GOSAT 2014 line list \cite{nikitin_gosat-2014_2015} (which was the basis of both HITRAN2016 and HITRAN2020 for this spectral region), the widely used empirical line list of Devi 2015 \cite{devi_self-_2015}, and the Nikitin 2017 analysis of WKLMC \cite{nikitin_analysis_2017}.  Significant differences exist between these line lists, complicated by correlation between parameters in these dense and complex spectra and the difficulty in identifying and assigning weaker transitions.  These complications lead to line lists that trade parameter accuracy for self-consistency, thus impeding efforts to refine line parameters.

        Recently, this problem has been examined in the methane tetradecad \textit{via} three separate sets of experiments focused on the strong $2\nu_3$ band of \textsuperscript{12}CH\textsubscript{4}, which represents $\approx{67\%}$ of the total summed intensity of the R-branch spectral region from 6010 cm$^{-1}$ to 6120 cm$^{-1}$.  Each experiment was tailored for optimal determination of particular spectroscopic parameters: saturation spectroscopy for transition frequencies, linear absorption spectroscopy in the Doppler-broadened domain for absolute transition intensities and ratios, and line shape and line mixing parameters determined in the pressure broadened domain. Transition frequencies for 53 transitions in the $2\nu_3$ R-branch were determined with kHz-level uncertainty through saturation spectroscopy by Votava \textit{et al.} \cite{votava_2022}. These highly accurate values were subsequently used as constraints in the determination of molecular line intensities in the Doppler-broadened domain through independent experiments employing CRDS at NIST and FTS at DLR as reported by Reed {\it et al.} \cite{reed_multilaboratory_2025}, which additionally replaced $\approx 2600$ weaker non-$2\nu_3$ lines from the Nikitin 2017 line list. Permille-level uncertainties were demonstrated (evaluated by relative agreement between the two independent measurements) for 64 strong transitions, covering the $2\nu_3$ R-branch from $m =$ (1--11).  These line intensities were employed by Yin \textit{et al.}\cite{yin_2025} in further CRDS studies involving elevated pressure of CH\textsubscript{4} broadened by N\textsubscript{2} to determine line shape and line mixing parameters. 

        This commonly shared set of physical constraints provides a consistent physical basis for the $2\nu_3$ band parameters.  The Nikitin 2017 line list \cite{nikitin_analysis_2017} was employed as the basis for the full line list from 5970--6250 cm$^{-1}$. The parameters of $\approx 2600$ lines are updated based on fits of the two-temperature (210 K and 298 K) FTS experiments at DLR, including the Q- and R- branches and a few P-branch manifolds.  The final line list required adding 105 transitions and removing 57 transitions from Nikitin 2017. For 238 transitions, the lower state energies, $E''$, were found to be incompatible with those of Nikitin 2017, and consequently these  values were updated in HITRAN2024. A comparison of the intensities in the HITRAN2020 and HITRAN2024 line lists can be found in Figure \ref{NIST DLR Overlap set}. Note that approximately 2200 additional transitions with  $S < 2\times 10^{-25}$ cm$^{-1}$/(molec$\times$cm$^{-2})$ are included from Nikitin 2017.

        \begin{figure}[!htbp]
        \includegraphics[width=0.9\textwidth]{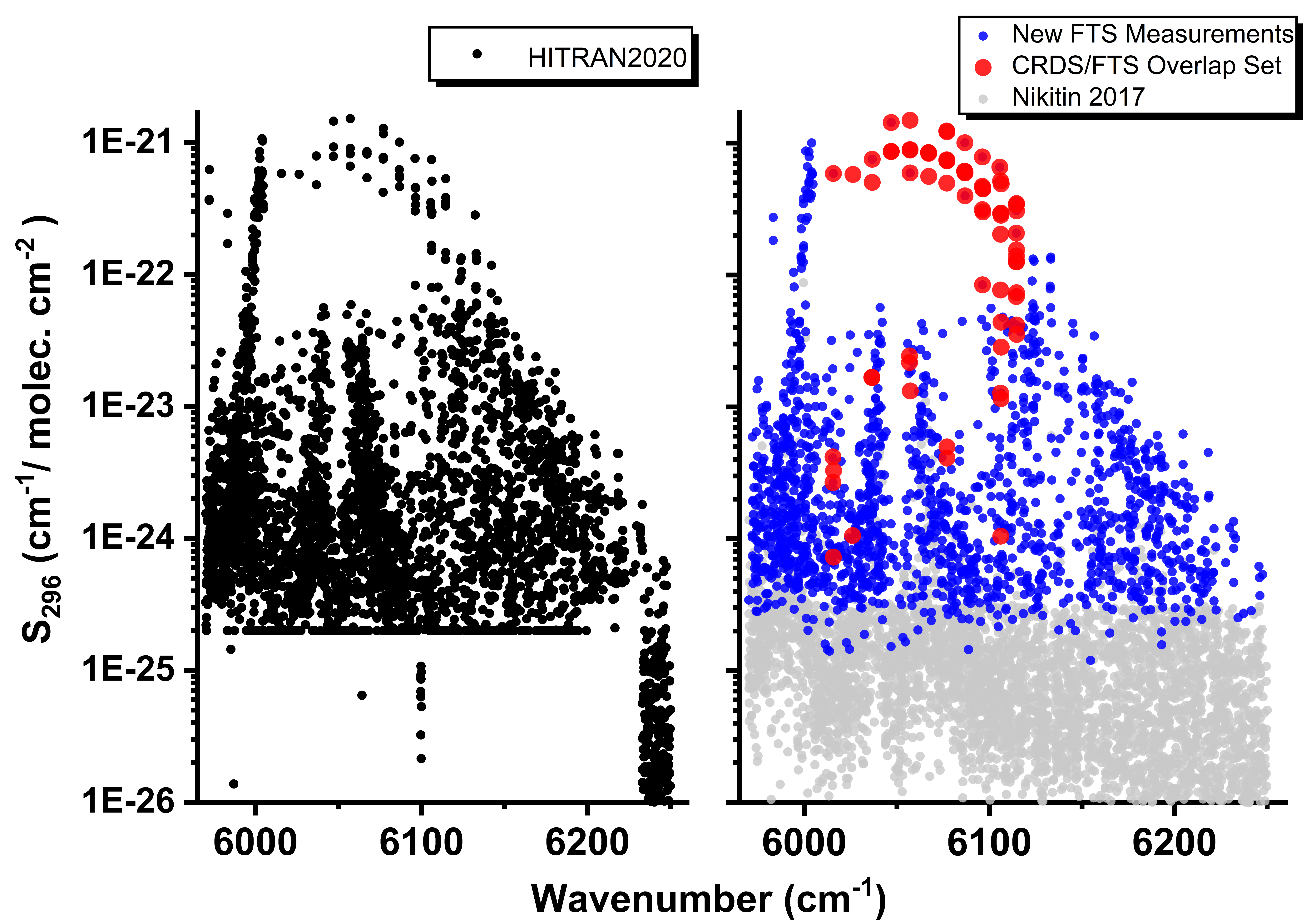}
        \caption{Comparison of the intensities for the HITRAN2020 (left panel) and HITRAN2024 line lists (right panel) in the \textsuperscript{12}CH\textsubscript{4} tetradecad, including the $2\nu_3$ band.  The overlap set includes transition intensities determined at NIST (CRDS) and DLR (FTS) in the Doppler-broadened domain.}
        \label{NIST DLR Overlap set}
        \end{figure}

        Significant improvements in fit quality over previously determined spectroscopic parameters are demonstrated from Doppler-broadened conditions to atmospherically relevant pressure-broadened cases.  The R(10) manifold is especially highly blended and the accuracy of the line intensities is significantly limited by fitting correlation.  Given the large observed differences between DLR and NIST in the R(10) manifold, the NIST parameters were selected and used to constrain the USTC lineshape determinations at elevated pressures.  Consequently, the resulting parameters reported for R(10) should be considered an effective parameter set.   

        Agreement with TheoReTS \cite{rey_theorets_2016} is excellent (at the several permille level) for the assigned $2\nu_3$ transition intensities, however, the lack of assignments in the experimental line list for many weaker (non-$2\nu_3$) transitions makes direct comparison difficult.  {\color{blue}Limited experimental signal-to-noise or fitting correlation requires the inclusion of many unassigned weak transitions in the experimental linelist to fully model the observed spectra.  These transitions likely represent the sum of a number of weaker transitions predicted by THeoReTS.}

        \subsubsection{LIPhy DAS measurements of \texorpdfstring{\textsuperscript{12}CH\textsubscript{3}D}{12CH3D}}
        In spite of the small magnitude of the CH\textsubscript{3}D/CH\textsubscript{4} amount ratio, (nominally $5\times 10^{-4}$ at natural abundance), CH\textsubscript{3}D has a large relative contribution {\color{blue} to the methane absorption} in the region (up to 70 \% near 6300 cm$^{-1}$ at 80 K) \cite{wang_high_2010}) and contributes greatly to the absorption in the 1.58 μm methane transparency window \cite{lu_ch3d_2011}. Nevertheless, the $3\nu _2$ band is not isolated but superimposed on a congested spectrum of lines involving other weaker bands which have to be characterized for a proper description of the absorption in the region. As described in \cite{brown_methane_2013}, since its 2012 edition, the HITRAN list was extended to the 6204-6510 cm$^{-1}$ region by including for the 6204--6394 cm$^{-1}$ interval, the empirical line list constructed in Lu \textit{et al.} \cite{lu_ch3d_2011} using diode laser absorption spectroscopy (DAS) of highly enriched CH\textsubscript{3}D at 296 K. This list of about 5500 lines included about 175 rovibrational assignments and more than 2700 empirical values of the lower state energy, $E''_\mathrm{emp}$, obtained from the ratios of intensities measured at 296 K and 80 K (2T-method). 
    
        The update of CH\textsubscript{3}D in this spectral region is based on the recent work of Ben Fathallah \textit{et al.} \cite{benfathallah_ch3d_2024} which extends that of \textit{Lu et al.} The region of the DAS recordings is increased to include the 6099--6530 cm$^{-1}$ region (11189 lines) and about 4800 $E''_\mathrm{emp}$ values were obtained by the 2T-method. Relying on the position and intensity agreements with the TheoReTS variational line list \cite{rey_accurate_2014,rey_theorets_2016} about 2890 transitions could be fully rovibrationally assigned to twenty bands, fifteen of them being newly reported. All the reported assignments were confirmed by Ground State Combination Difference (GSCD) relations i.e. all the upper state energies (about 1370 in total) have coinciding determinations through several transitions (up to 8). The intensity sum of the transitions assigned between 6190 and 6530 cm$^{-1}$ represents 76.9 \% of the total experimental intensities at 296 K.

        {\color{blue}\subsubsection{HITRAN update in the 5900--6530 \texorpdfstring{cm$^{-1}$}{cm-1} region}}
        \begin{figure}[!htbp]
            \includegraphics[width = \linewidth]{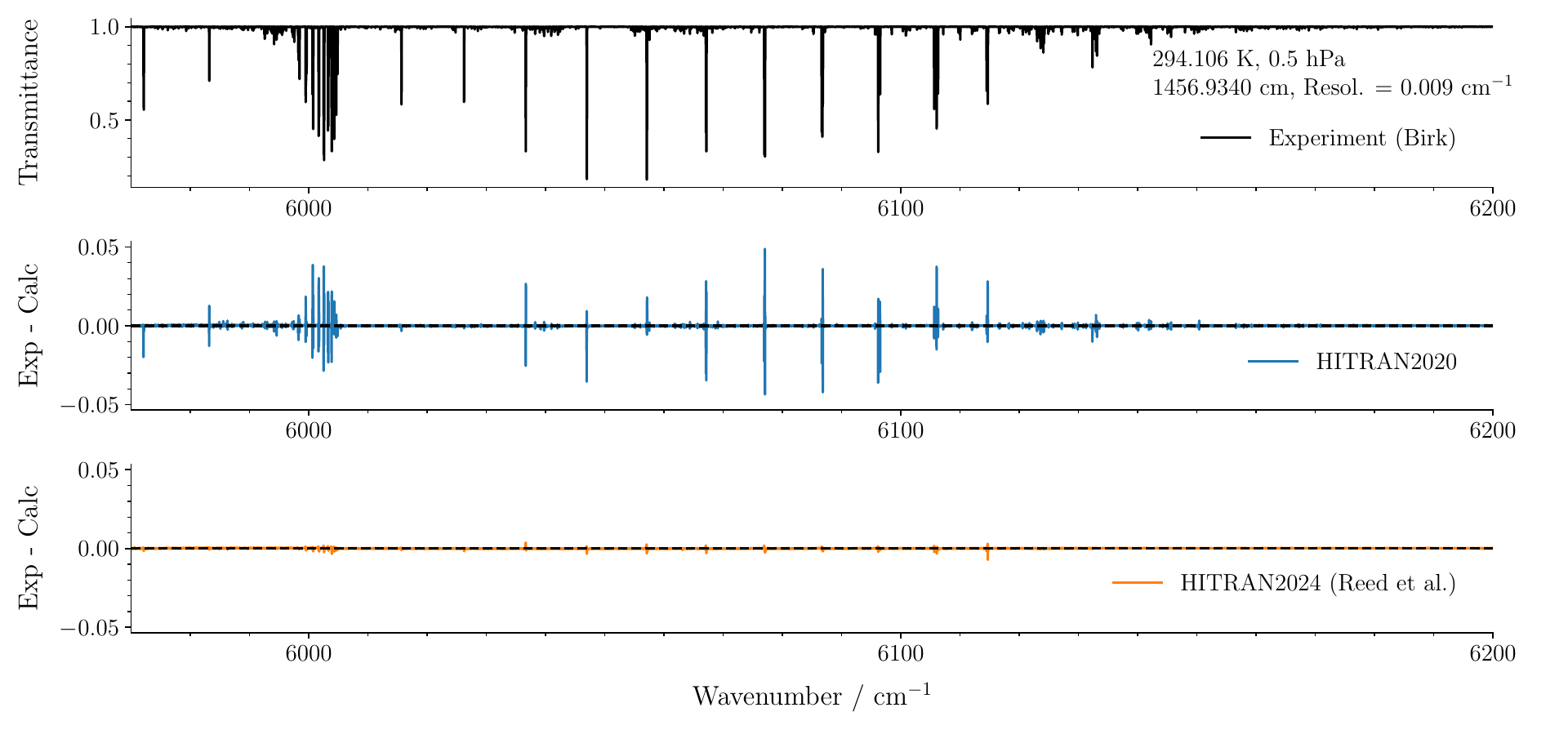}
            \caption{Residuals obtained at 294 K and 0.5 hPa between the experiment from Ref. \cite{birk_measurement_2025} and Voigt calculations from the line lists used for the HITRAN2024 update.}
            \label{fig:birk_tetradecad_sources}
        \end{figure}
        The DLR/NIST line list \cite{reed_multilaboratory_2025,yin_2025} based on Nikitin 2017 \cite{nikitin_analysis_2017} replaced the GOSAT 2014 \textsuperscript{12}CH\textsubscript{4} transition wavenumbers, and intensities in their region, for more than 2500 transitions. In several manifolds, better residuals were achieved with fewer lines than previously included in HITRAN2020 (Fig. \ref{fig:birk_tetradecad_sources}). As such, some lines were removed to accommodate for these changes as well. The new pressure-induced parameters were used to replace the previous values, as discussed below in Sec. \ref{sec:pressure_induced}.
        
        The new list from Ben Fathallah et al. \cite{benfathallah_ch3d_2024} {\color{blue}that} uses their experimental \textsuperscript{12}CH\textsubscript{3}D, and variational calculations if not available, {\color{blue} was included} in HITRAN2024. More than 5800 new transitions supplement the lower wavenumber range of the previous data. These data comprise 4800 lines providing updates on the positions, intensities, {\color{blue}and} energy levels, as well as new assignments.

    {\color{blue}\subsection{6770--7570 \texorpdfstring{cm$^{-1}$}{cm-1}}\label{sec:6770-7570}}
        \subsubsection{PLDL Dual frequency comb}
        
        Malarich \textit{et al.} \cite{malarich_dual_2021} present new broadband, high-resolution dual frequency comb spectroscopy (DCS) of pure methane up to 1000 K and 300 Torr in the methane icosad (6770--7570 cm$^{-1}$). The dual comb spectrometer spans 870 cm\textsuperscript{-1} with 0.00667 cm\textsuperscript{-1} point spacing and a wavenumber uncertainty of 10\textsuperscript{-5} cm\textsuperscript{-1}. Light from the spectrometer was passed twice through a 46-cm-long optical cell (92 cm total pathlength) placed in a uniform three-zone furnace. 

        Thirteen total spectra were analyzed at five temperatures from 296-1000 K and 18-300 Torr. Simulations in the icosad region with HITRAN2020 show good agreement with the measured spectra at room temperature, but exhibit significant mismatches at higher temperatures. This is not surprising given that the empirically based HITRAN2020 icosad line list originated from spectra measured below 280 K (WKLMC \cite{campargue_2013}). The new high-temperature spectra were used to update many HITRAN2020 lower-state energies, line positions and self shifts, self widths, and power-law temperature dependence exponents for self width and shift. 
        
        Much of the high-temperature discrepancy between HITRAN2020 and the DCS data was due to methane absorption features that had been given a default value of lower-state energy (such as $E"\textsubscript{HITRAN}=999$ cm$^{-1}$). Using multispectrum fitting to the high-temperature spectra, 4283 lower-state energies are assigned, which had previously been given this default value. Existing lower-state energies were also updated for 92 lines, and the database adds 293 high-temperature lines that improve the database performance above 300 K. The method for these updates is described in detail in \cite{malarich_dual_2021}. Fig.8 in Ref. \cite{malarich_dual_2021} shows the distribution of new lower-state energy values.


        
        The lowest-pressure methane spectrum shows a statistically significant different non-zero shift in three different spectral regions (see Fig.5 of Ref. \cite{malarich_dual_2021}). The sudden change in $\Delta$\textsubscript{fit} at 7050 and 7400 cm\textsuperscript{-1} appears in all methane spectra but not in the water vapor measurement, so it can be concluded that this was a HITRAN2020 vacuum-position error and not an artifact. Therefore, a vacuum position correction $\delta_{\nu_0}$ was applied for each of the three spectral regions.

        A similar approach was followed to determine band-wide average adjustments to self broadening (+7\%), and to offer new power-law temperature scaling exponents of $n_\gamma=0.85$ and $n_\delta=0.58$. Altogether, these updates greatly reduce the model-measurement residuals for the icosad region across a large array of temperature and pressure conditions, as shown, for example at 623 K and 300 Torr in Fig. \ref{fig:IcosadComparison}. 

         \begin{figure}[!htbp]
            \centering
            \includegraphics[width = 1\textwidth]{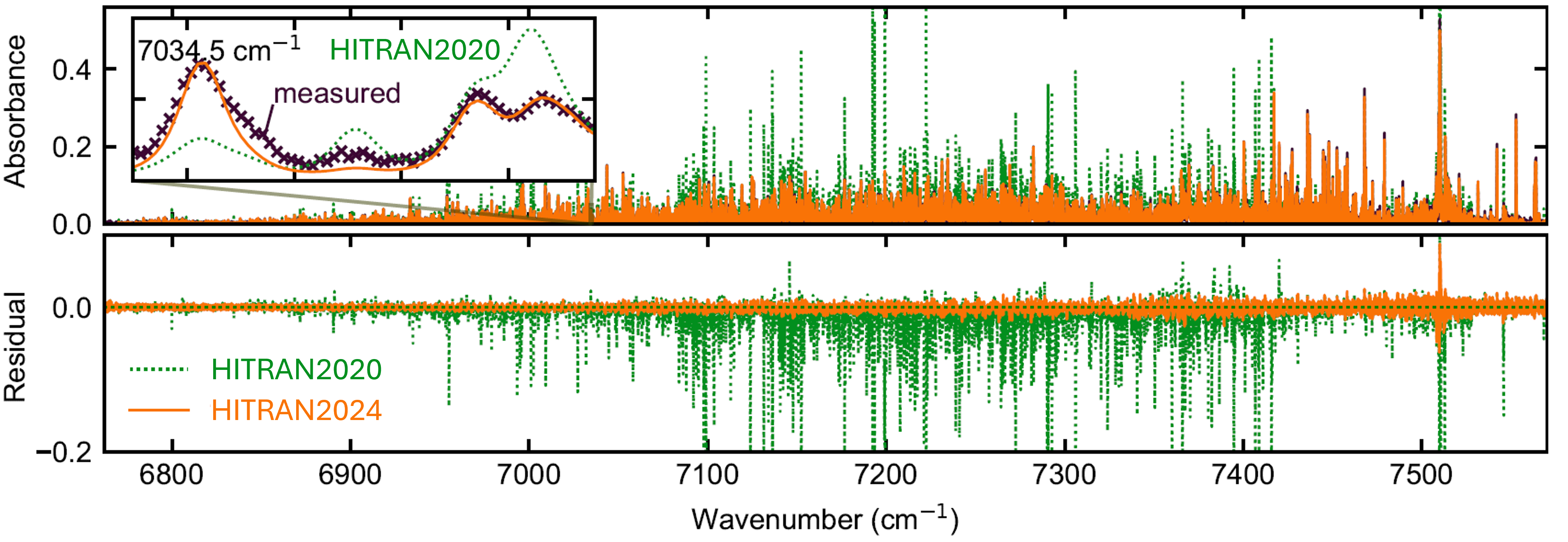}
            \caption{Representative high-temperature DCS spectrum, at 623 Kelvin, 300 Torr (brown crosses) \cite{malarich_dual_2021}. Much of the HITRAN2020 residual (green dotted traces) is due to features that have the default $E"\textsubscript{HITRAN2020}=999.99$ cm$^{-1}$. The net negative residual indicates that $E"\textsubscript{HITRAN2020}$ is too large. Orange trace shows final model with updates described here and in \cite{malarich_dual_2021}.}
            \label{fig:IcosadComparison}
        \end{figure}

        {\color{blue}\subsubsection{HITRAN update in the 6770--7570 \texorpdfstring{cm$^{-1}$}{cm-1} region}}
        The new PLDL list \cite{malarich_dual_2021} mostly replaces the \textsuperscript{12}CH\textsubscript{4} transition wavenumbers and some lower energy levels of HITRAN2020 while keeping the old intensities. More than 22,000 transitions were updated using their data, and about 300 new lines were added. {\color{blue} The improved residuals between the measurement and updated database are shown in Fig. \ref{fig:IcosadComparison}.} Similarly to the DLR/NIST/USTC measurements, the new pressure-induced parameters were used as described in Section \ref{sec:pressure_induced}.

    {\color{blue}\subsection{7600--7920 \texorpdfstring{cm$^{-1}$}{cm-1}}\label{sec:7600-7020}}
        \subsubsection{IAO \& LIPhy CRDS measurements}
        An extended assignment and analysis of \textsuperscript{12}CH\textsubscript{4} lines in the 7606--7919 cm$^{-1}$ region was recently reported by Nikitin \textit{et al.} \cite{nikitin_first_2024}. The experimental methane spectra in this range recorded with CRDS technique in Grenoble  \cite{mondelain_2011} at 80 K and 296 K correspond to the high-energy part of the WKLMC empirical line list  \cite{campargue_2013}.
       Some lines in this range belong to the upper edge of the Icosad band system previously considered in \cite{rey_2015,rey_2016}, but most of the other transitions have been included in HITRAN2020 without assignments.
    
        Quantum assignments for 1382 transitions pointing to 33 vibrational sub-levels of the \textsuperscript{12}CH\textsubscript{4} Triacontad system were obtained using a combined approach based on variational \textit{ab initio} predictions \cite{rey_2013,rey_new_2018} with a subsequent empirical optimization of the effective spectroscopic models derived from the potential and dipole moment energy surfaces as described in \cite{tyuterev_2022,rey_2022}. They belong to three weak Triacontad bands, $6\nu_4$, $\nu_2$+$5\nu_4$, $\nu_1$+$4\nu_4$, falling in the methane transparency window near 1.28 μm. The positions of these vibrational bands are in good agreement with the values obtained from the \textit{ab initio} potential energy surface \cite{nikitin_var_2018,nikitin_pes_2025}.
        The root-mean-square fit deviation for the line positions of newly assigned transitions was 0.006 cm$^{-1}$. The intensities of the line list for these bands span the range $10^{-25} $ to $9 \times 10^{-30}$ cm/molecule.
        The full statistical results of the analyses are given in Table 1 of  \cite{nikitin_first_2024} and an overview plot of the \textsuperscript{12}CH\textsubscript{4} experimental line lists near 1.28 μm is shown in Figure 4 of \cite{nikitin_first_2024}.

        {\color{blue}\subsubsection{HITRAN update in the 7600--7920 \texorpdfstring{cm$^{-1}$}{cm-1} region}}
        The line list provided by Nikitin \textit{et al.} \cite{nikitin_first_2024} completes the \textsuperscript{12}CH\textsubscript{4} main band in the lower part of the triacontad. More than 400 new lines were added, and more than 50 were updated on their positions, intensities and lower state energy levels. Another major benefit of the update in that region comes with the addition of rotational assignments, which were previously very scarce and are now mostly complete for these bands. With $E_\mathrm{Lower}$, this allows for modeling in a wider range of temperatures and gives better estimation of line-shape parameters.
        
    {\color{blue}\subsection{10800-14000 \texorpdfstring{cm$^{-1}$}{cm-1}}\label{sec:10800-14000}}
        \subsubsection{LIPhy Kitt peak FTS and CRDS measurements 10802--13922 \texorpdfstring{cm$^{-1}$}{cm-1}}
        \begin{figure}[!htbp]
            \centering
            \includegraphics[width = 0.9\textwidth]{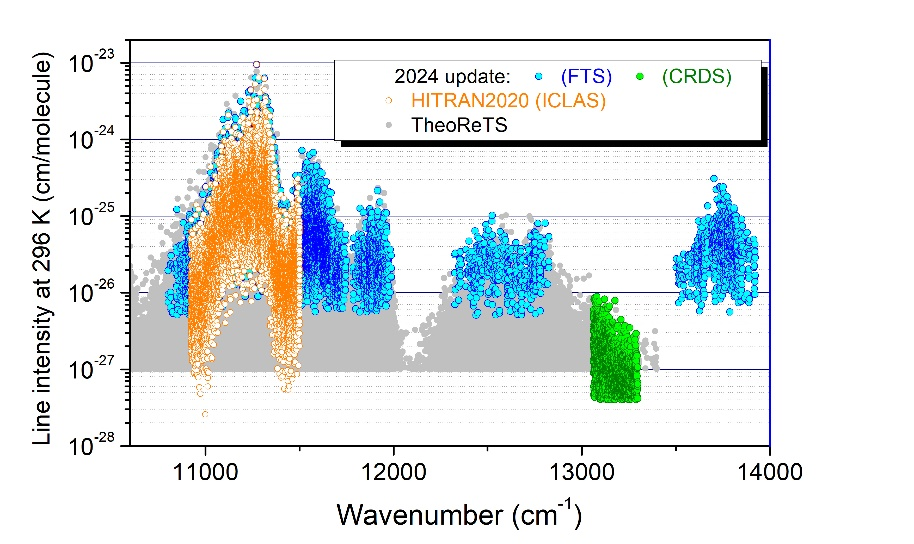}
            \caption{Update of the methane line list above 10800 cm$^{-1}$. The HITRAN2020 list (orange circles) was completed by new FTS and CRDS measurements (blue and green circles, respectively). The TheoReTS variational list \cite{rey_new_2018} (grey circles) is plotted as background below 13500 cm$^{-1}$.}
            \label{fig:intensity_10800-14000}
        \end{figure}
        In Ref. \cite{campargue_high_2023}, empirical line lists at room temperature were retrieved from a long-path Kitt Peak FTS spectrum between 10802 and 13922 cm$^{-1}$ (about 12800 lines) and from a high-sensitivity CRDS spectrum in the 13060--13300 cm$^{-1}$ interval (about 2650 lines). Above 10800 cm$^{-1}$, the previous HITRAN2020 dataset was limited to the 10923--11502 cm$^{-1}$ interval where measurements obtained by intracavity laser absorption spectroscopy (ICLAS) were used as source (see Brown \textit{et al.} \cite{brown_methane_2013}). The FTS results of Ref. \cite{campargue_high_2023} were found to be in good agreement with the HITRAN2020 dataset.
        
        At these wavenumbers, the methane spectrum is especially highly congested even at room temperature, and, according to Ref. \cite{campargue_high_2023}, part of the absorption shows up as a ``quasi-continuum'', which should be added to the absorption lines to fully account for the methane absorption in the region. The ``quasi-continuum'' represents about one third of the total absorption in the 11000--11400 cm$^{-1}$ region. Note that the intensity sum of the lines and of the continuum was found to be in good agreement with the intensity predictions of the TheoReTS list \cite{rey_new_2018}. {\color{blue} In this context, one can use the weak lines from the HITEMP database \cite{10.3847/1538-4365/ab7a1a} (which is largely based on TheoReTS, especially in this region)}.

       {\color{blue}\subsubsection{HITRAN update in the 10800--14000 \texorpdfstring{cm$^{-1}$}{cm-1} region}}
       This new list from Campargue \textit{et al.} \cite{campargue_high_2023} completes the update below 10923 cm$^{-1}$ and above 11502 cm$^{-1}$: regions that were previously empty (see Fig. \ref{fig:intensity_10800-14000}), with an additional 5000 FTS and 1800 CRDS new transitions, and excluding the quasi-continuum, which contributes to the total absorption and will need to be accounted for in the future. These lines are new to HITRAN. As a consequence of being completely empirical, the new lines were attributed to the main isotopologue by default. This issue will have to be addressed in a future HITRAN update.


    \subsection{Overview}
    In total, about 125,000 lines have either been modified or added from the lists described above to the database, increasing the total number of lines from 446,000 in HITRAN2020 to 471,000 now.
    \begin{figure}
        \includegraphics[width = \textwidth]{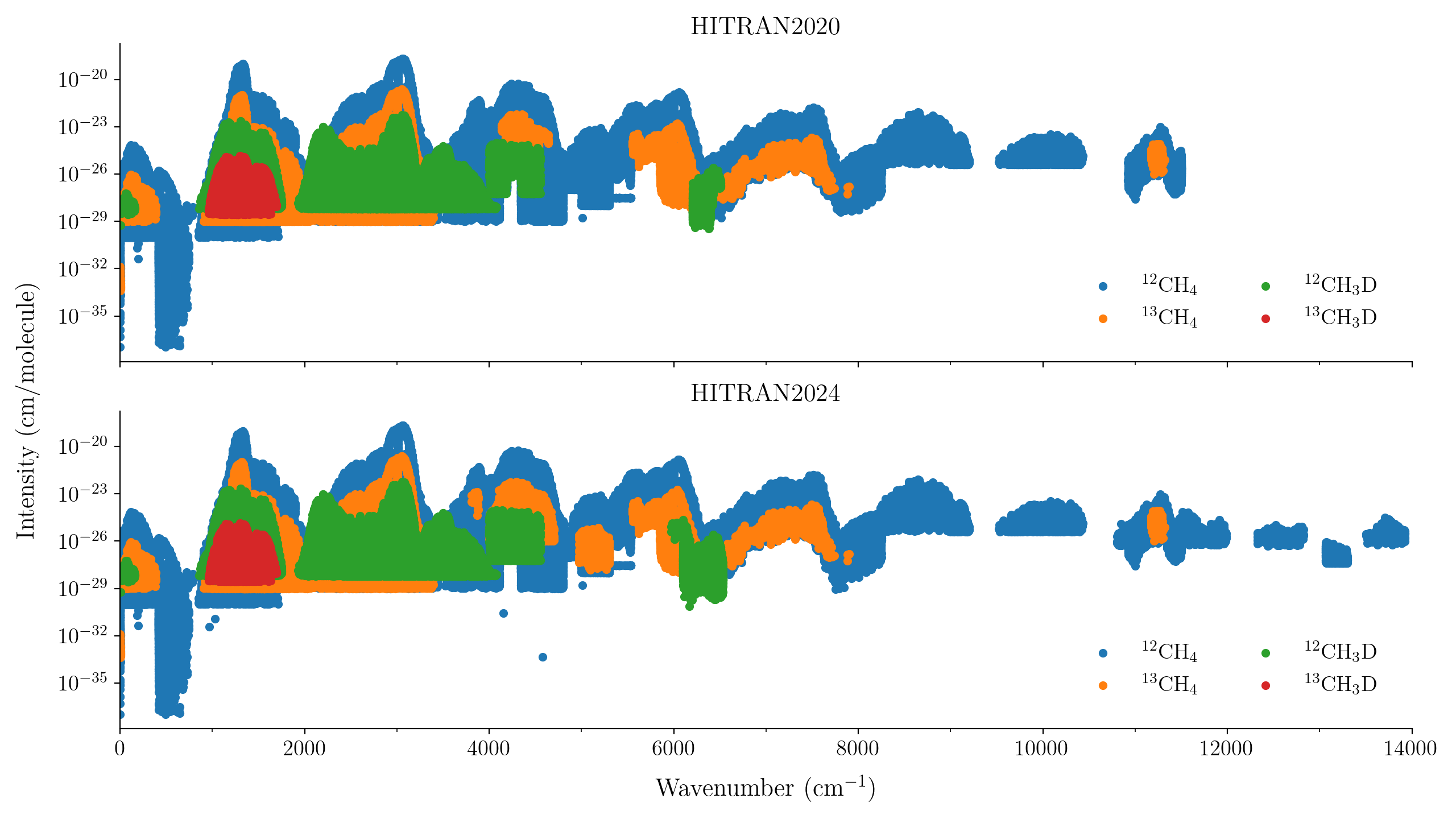}
        \caption{Line list overviews: Intensities of each line included in HITRAN2020 (upper panel) and HITRAN2024 (lower panel) for every isotopologues of methane}
        \label{fig:hitran_intensities}
    \end{figure}
    The extent of these additions can be visualized in Fig. \ref{fig:hitran_intensities}, where we see that \textsuperscript{12}CH\textsubscript{4} lines have been expanded up to 14000 cm$^{-1}$ and, with the other isotopologues, completed some of the gaps. Additionally, 16,000 HITRAN2020 lines were removed; however, updated unassigned lines, or those with an incorrect assignment format in HITRAN2020, are counted as removed and added. These constitute the majority of removed lines, because they could not be automatically matched in other line lists. Some of them now have complete or at least partial assignments.
    
    \begin{table}[htbp]
        \caption{Source of the new line parameters in the HITRAN2024 update}
        \label{table:origin_parameters}
        \begin{tabular}{llrrr}
\toprule
Line list & Isotopologue & \makecell*[{{p{14ex}}}]{Transition wavenumbers} & Intensities & Region covered (cm$^{-1}$)\\
\midrule
Nikitin \textit{et al.}, 2025 \cite{nikitin_improved_2024} & \textsuperscript{12}CH\textsubscript{4} & 1880 & 1610 & 900 -- 1050\\
 & \textsuperscript{13}CH\textsubscript{4} & 60 & 40 & \\
 & \textsuperscript{12}CH\textsubscript{3}D & 500 & 460 & \\
 & \textsuperscript{13}CH\textsubscript{3}D & 60 & 60 & \\
\midrule
ExoMol \cite{yurchenko_exomol_2024} & \textsuperscript{12}CH\textsubscript{4} & 240 & 140 & 2503 -- 4177\\
\midrule
MeCaSDa \cite{RICHARD2024109127} & \textsuperscript{12}CH\textsubscript{4} & 70 & 30 & 2786 -- 3910\\
 & \textsuperscript{13}CH\textsubscript{4} & 30 & 30 & \\
\midrule
Sung \textit{et al.} \cite{sung_new_2024} & \textsuperscript{13}CH\textsubscript{4} & 4710 & 4710 & 4000 -- 4700\\
\midrule
Nikitin \textit{et al.}, 2013 \cite{nikitin_preliminary_2013} & \textsuperscript{12}CH\textsubscript{3}D & 1530 & 1510 & 4001 -- 4554\\
\midrule
Rodina \textit{et al.}, 2021 \cite{rodina_improved_2021} & \textsuperscript{12}CH\textsubscript{4} & 4930 & 4230 & 4100 -- 4190\\
 & \textsuperscript{13}CH\textsubscript{4} & 430 & 430 & \\
 & \textsuperscript{12}CH\textsubscript{3}D & 10 & 10 & \\
\midrule
atm.161 \cite{toon_atmospheric_2022} & \textsuperscript{12}CH\textsubscript{4} & 22320 & 23450 & 4340 -- 4899\\
 & \textsuperscript{13}CH\textsubscript{4} & 1710 & 1700 & \\
 & \textsuperscript{12}CH\textsubscript{3}D & 500 & 500 & \\
\midrule
Starikova \textit{et al.}, 2024 \cite{starikova_assignment_2024} & \textsuperscript{13}CH\textsubscript{4} & 1640 & 1640 & 4970 -- 5300\\
\midrule
Reed \textit{et al.}, 2025 \cite{reed_multilaboratory_2025} & \textsuperscript{12}CH\textsubscript{4} & 7150 & 7270 & 6016 -- 6115\\
 & \textsuperscript{13}CH\textsubscript{4} & 650 & 660 & \\
 & \textsuperscript{12}CH\textsubscript{3}D & 60 & 60 & \\
\midrule
Ben Fathallah \textit{et al.}, 2024 \cite{benfathallah_ch3d_2024} & \textsuperscript{12}CH\textsubscript{3}D & 10630 & 5850 & 6098 -- 6530\\
\midrule
Malarich \textit{et al.}, 2021 \cite{malarich_dual_2021} & \textsuperscript{12}CH\textsubscript{4} & 22320 & 270 & 6770 -- 7570\\
\midrule
Nikitin \textit{et al.}, 2024 \cite{nikitin_first_2024} & \textsuperscript{12}CH\textsubscript{4} & 480 & 480 & 7629 -- 7919\\
\midrule
Campargue \textit{et al.}, 2023 \cite{campargue_high_2023} & \textsuperscript{12}CH\textsubscript{4} & 6780 & 6780 & 10803 -- 13923\\
\bottomrule
\end{tabular}

    \end{table}
    In general, more transition wavenumbers were taken from the new data than line intensities (see Table \ref{table:origin_parameters}), which stems from our criteria for replacing the existing parameters. From HITRAN2020, about 51,000 of the transition wavenumbers were replaced, with 18,000 for the intensities, 11,000 for the energy levels, and 3300 for the assignments. Given the general lower accuracy of the intensities provided, a larger difference between the HITRAN2020 values and the new data was required before updating the value. In most cases, uncertainties were not even reported. The SNR is another contributing factor to consider when most residuals at room temperature and natural isotopic abundance do not always exhibit an obvious improvement from the new lists, especially for weaker lines.

\section{Pressure-induced line-shape parameters}\label{sec:pressure_induced}

    \subsection{USTC, NIST \& DLR measurements}
    \begin{figure}
        \includegraphics[width = \textwidth]{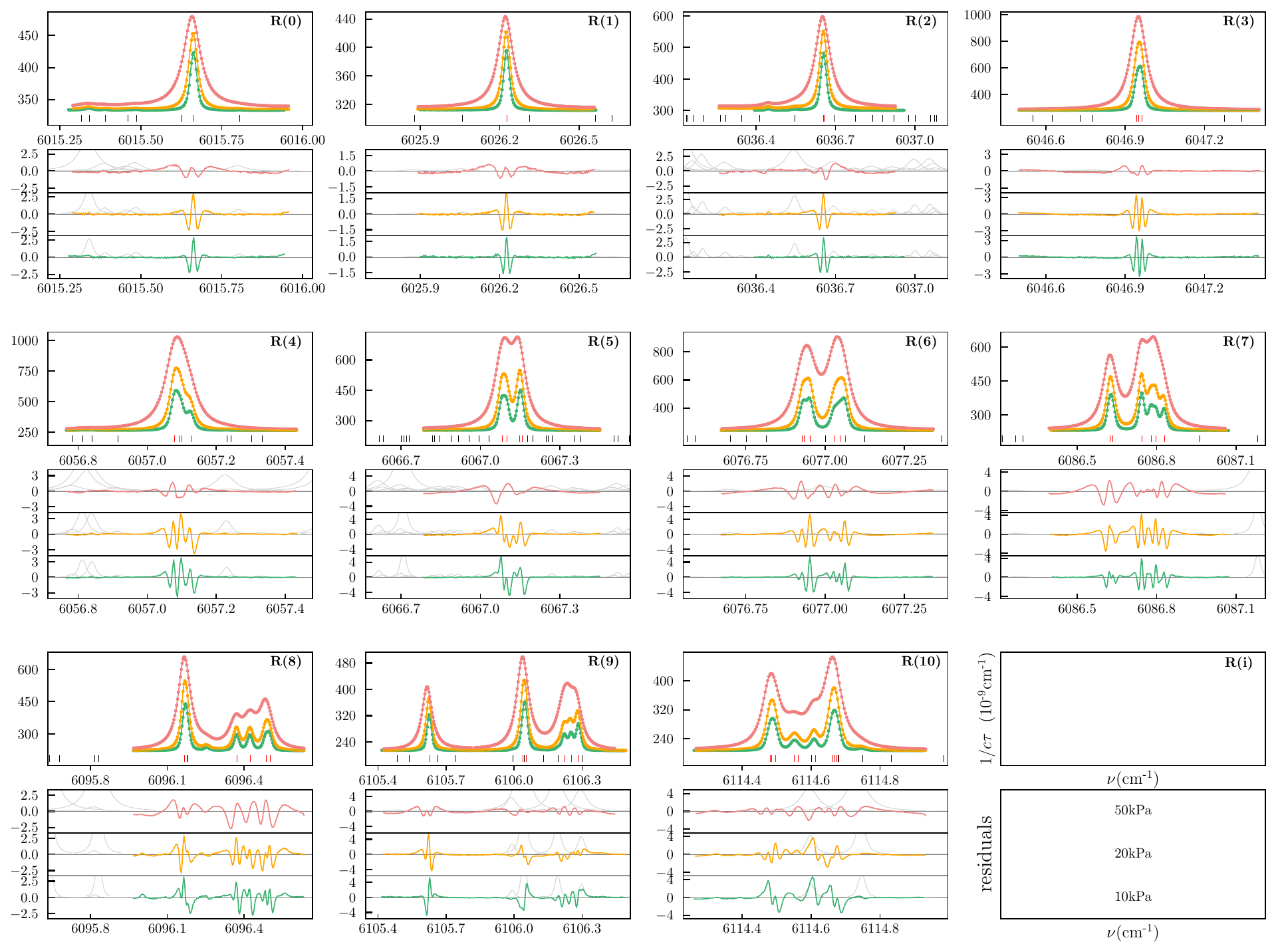}
        \caption{An overview of CRDS spectra and global fit results for the R(0)-R(10) transition with LMVP. }
        \label{fig:Yan_hitran}
    \end{figure}
    An extensive set of high-precision evaluations of the methane absorption spectrum at 1.64 μm, specifically aimed at improving the precision of spectral line profile parameters, was made at the University of Science and Technology of China (USTC) in collaboration with the US National Institute of Standards and Technology (NIST) and the German Aerospace Center (DLR). This work highlights a newly built cavity ring-down spectroscopy system characterized by exceptional thermal stability and accurate pressure calibration. This device enabled the acquisition of methane spectral data within a nitrogen environment, focusing on the R(0) to R(10) regions of the 2$\nu_3$ band of \textsuperscript{12}CH\textsubscript{4} at three distinct pressure levels spanning from 10 kPa to 50 kPa.
    
    For transitions in the $2\nu_3$ band of methane, HITRAN2020 only covers the parameters of Voigt profile (VP), i.e., pressure broadening ($\gamma_0$) and pressure shifting ($\delta_0$). Yet, to achieve greater precision in atmospheric methane simulations, more intricate line shapes and precise parameters are needed. In this regard, we employed the Speed-Dependent Voigt profile (SDVP) corrected for first-order Line-Mixing (LM), abbreviated as LMSDVP, as our base model to analyze the CRDS data. SDVP is a simplification of the pCqSDHC(HT) profile discussed in Ngo~\textit{et al.}\cite{NGO2013JQSRT}, which is more accurate than the VP owing to the inclusion of the speed-dependent effect. The involvement of numerous fitting parameters in the HT profile, compounded by the intricate nature of methane spectra, can result in closely numerically correlated spectral line parameters. This can impede the convergence of fitting solutions when analyzing the actual spectra or cause the derived physical parameters to be less physical. Therefore, we did not use the HT profile.  Indeed, the SDVP is adequate for most endeavors of spectral analysis and simulations, but additional treatment is needed due to the high degeneracy of methane states. Thus, we introduced the first-order Rosenkranz parameter, $Y_n$, to account for line mixing effects~\cite{Rosenkranz1975}. From a computational perspective, $Y_n$ exhibits a strong correlation with line intensity $S$. Consequently, an independent measurement of line intensity is crucial for acquiring dependable line-mixing parameters. Together with the intensity measurements by NIST and DLR, a global-fit procedure was developed based on the LMSDV model. We also converted the measured nitrogen-broadening coefficients to air-broadening coefficients using the formula $\gamma_0^\text{air} = 0.79\gamma_0^{\mathrm{N}_2} + 0.21\gamma_0^{\mathrm{O}_2}$, and utilizing polynomial parameters for $\gamma_0^{\mathrm{O}_2}$ as provided by Lyulin~\textit{et al.}~\cite{Lyulin2009JQSRT}. The nitrogen pressure shifting coefficients were similarly scaled as $\delta_0^\text{air} = 0.79\delta_0^{\mathrm{N}_2} + 0.21\delta_0^{\mathrm{O}_2}$, again utilizing parameters for $\delta_0^{\mathrm{O}_2}$ as provided by Lyulin~\textit{et al.}~\cite{Lyulin2009JQSRT}. Values of $Y_n^{\mathrm{N\textsubscript{2}}}$ for the measured CH\textsubscript{4}--N\textsubscript{2} system are converted for CH\textsubscript{4}--air through multiplication by 0.985, as shown in Tran \textit{et al.} for the $2\nu_3$ band \cite{Tran2010}. {\color{blue}The speed-dependent Voigt parameters were incorporated in HITRAN2024 wherever available.} 

    Many weaker transitions from 5970--6250 cm$^{-1}$, outside the NIST/DLR/USTC $2\nu_3$ overlap set, were updated with Voigt lineshape parameters ($\gamma_0, \delta_0$) recently determined from ambient temperature (nominally 298 K) FTS measurements performed by Birk \textit{et al.} at DLR.  The initial line list was the Nikitin 2017 line list \cite{nikitin_analysis_2017}, which was based on the empirical WKLMC line list, updated with the line intensity parameters described in Reed \textit{et al.} \cite{reed_multilaboratory_2025} and the line positions of Votava \textit{et al.} \cite{votava_2022}, where available.  Improved Voigt parameters are reported for approximately 1650 transitions with $S>3 \times 10^{-23}$ cm$^{-1}$/(molec$\times$cm$^{-2}$) outside the $2\nu_3$ band.  Lineshape parameters for $2\nu_3$ transitions were floated in the fit; however, the values adopted in HITRAN2024 are those determined by USTC above.

    \subsection{\texorpdfstring{\textsuperscript{12}CH\textsubscript{4} \& \textsuperscript{13}CH\textsubscript{4}}{12CH4 \& 13CH4}}
    Broadening parameters ($\gamma _0$) were gathered from all post-year 2000 sources \cite{antony_n2_2008,bertin_co2_2024,clark_difference_2004,devi_self-_2015,devi_spectral_2016,devi_multispectrum_2018,es-sebbar_intensities_2014,es-sebbar_linestrengths_2021,farji_airinduced_2021,gabard_line_2010,gharib-nezhad_h2induced_2019,ghysels_spectroscopy_2011,ghysels_temperature_2014,grigoriev_estimation_2001,lyulin_measurements_2014,ma_temperature_2016,manne_determination_2017,mondelain_line_2005,mondelain_measurement_2007,pine_speeddependent_2019,pine_speeddependent_2000,pine_multispectrum_2003,richard_self_2023} that we could find, and separated by collision partners, transition symmetries, and ranking index. Unlike for other rotors, only the ranking index, $\alpha $, lets us differentiate between lines of the same manifold and symmetry, which simply relates to the energy level instead of a motion. Because of this, the parameters do not exhibit any clear pattern with $m$ for each index (see Fig. \ref{fig:broad_alpha_1}). Thus, separation was only used to isolate the most intense lines, which are found in $\alpha = 1$.
    \begin{figure}[!htbp]
        \centering
        \includegraphics[width=\linewidth]{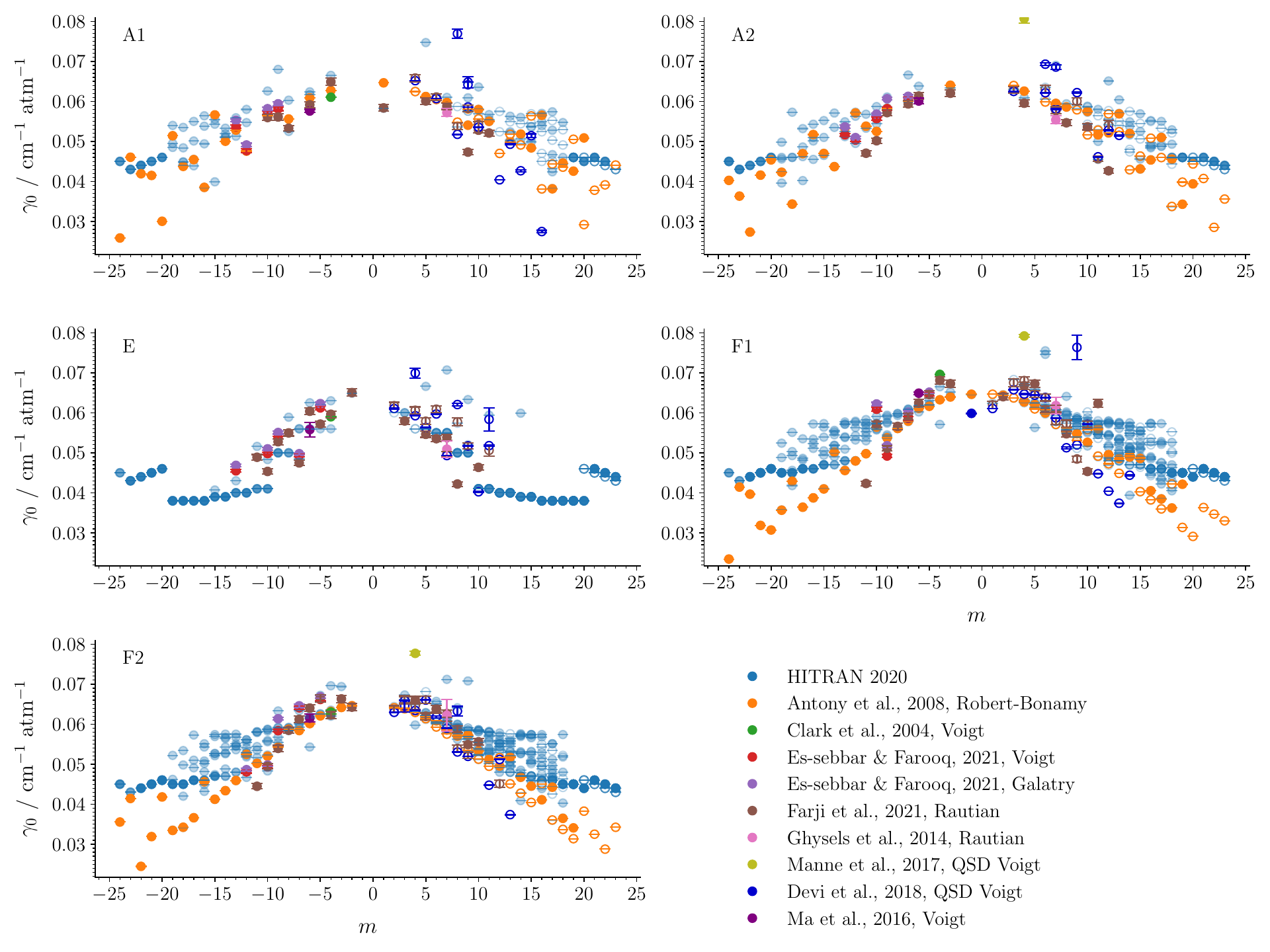}
        \caption{Air-broadening of methane for the first ranking index, $\alpha '' = 1$, displayed from the many sources available in literature. The HITRAN2020 data (in lighter blue) is shown with transparency. The more opaque points show multiple transitions having the same value.}
        \label{fig:broad_alpha_1}
    \end{figure}

    The different line profiles used in these studies impact the value of the retrieved coefficients by a small but non-negligible amount. For consistency with the parameters already present in HITRAN, we chose to focus on the Voigt profile. Separating each profile would greatly reduce the amount of data per fit; therefore, we scaled the respective non-Voigt parameters {\color{blue} by a constant value to match the Voigt broadening. The constant scaling was calculated by taking the ratio of Voigt to non-Voigt broadening for each transition and averaging the result.} This approach could only be performed in cases where transitions were fit with both non-Voigt and Voigt profiles \textemdash thus not all studies could be included. To further increase the amount of available data, the CH\textsubscript{4}--N\textsubscript{2} pressure broadening parameters were combined with those of --air, and scaled by a factor of 0.988 taken from Pine \& Gabard \cite{pine_speeddependent_2000}. This approach was previously used in other studies \cite{mondelain_measurement_2007} and agrees very well with other measurements (see Fig. \ref{fig:pade_fit_pr}). Finally, thanks to the weak branch dependence, broadening parameters for the $P$, $Q$, and $R$ branches were combined. Ideally, each branch should be separate; however, the broadening measurements in methane are too uncertain for the difference to be reliably quantified.
    
    Using a few assumptions, we can constrain the Padé approximant function to ensure convergence of the fit. First, because pressure broadening is positive-valued, we will force every parameter to be positive for simplicity. This constraint also prevents discontinuities in the curve by removing the possibility of division by zero. Second, based on measurements in various species, not only methane, we can assume that in the $\lim _{J \to \infty}$,  $\gamma _0$ is equal to zero. For this reason, it is preferable to have a polynomial of greater order in the divisor. We found that a {\color{blue}[2/3]} function of the form
    \begin{align}
        \gamma _0 = \dfrac{a_0 + a_1 m + a_2 m^2}{1 + b_1 m + b_2 m^2 + b_3 m^3}
    \end{align}
    works well for our level of precision. As observed in Fig. \ref{fig:broad_alpha_1}, even a single symmetry and ranking index does not lead to a smooth looking curve. Some transitions give systematically stronger or weaker parameters. Good examples of that behavior are the $J'' = 4$, $\alpha = 1$ transitions of the $F_2$ rotational symmetry, both in $P$ and $R$ branches, with parameters being weaker than those of their surrounding lines. This behavior appears in most experiments and is probably not an experimental artifact. Because these values deteriorate the fit, such points are removed beforehand.
    
    \begin{figure}[!htbp]
        \centering
        \includegraphics[width=0.5\linewidth]{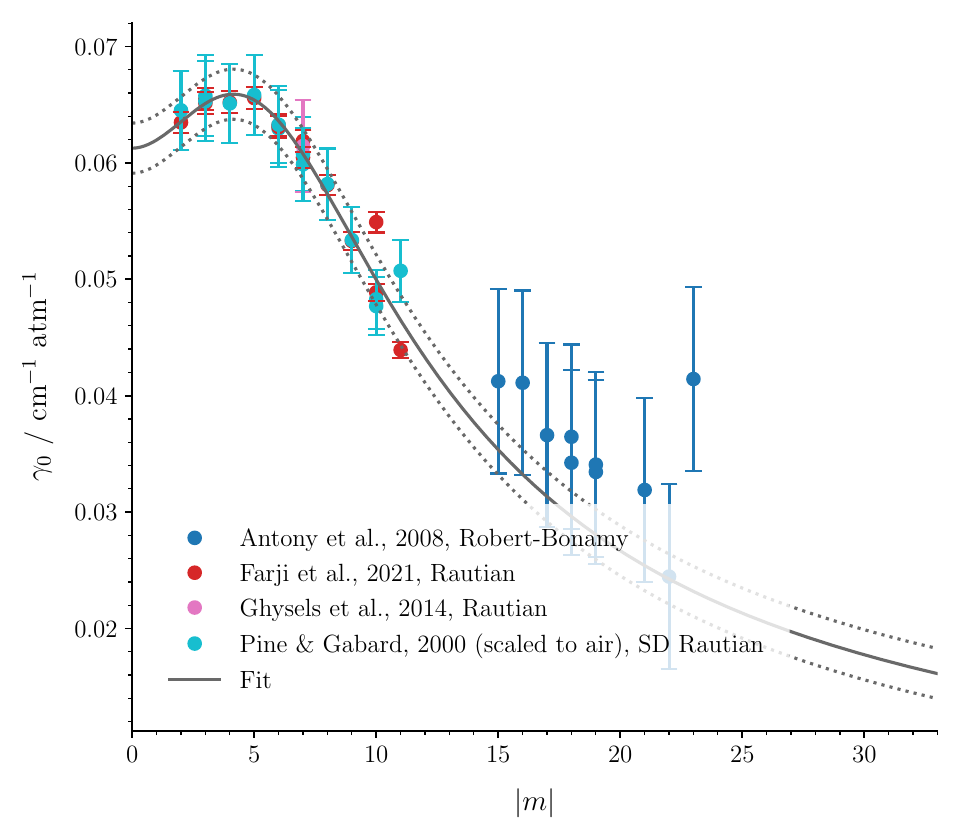}
        \caption{Data used for the fit, and resulting Padé approximant in the $F_2$ symmetry, combining $P$ and $R$ branches}
        \label{fig:pade_fit_pr}
    \end{figure}
    \begin{figure}[!htbp]
        \centering
        \includegraphics[width=0.5\linewidth]{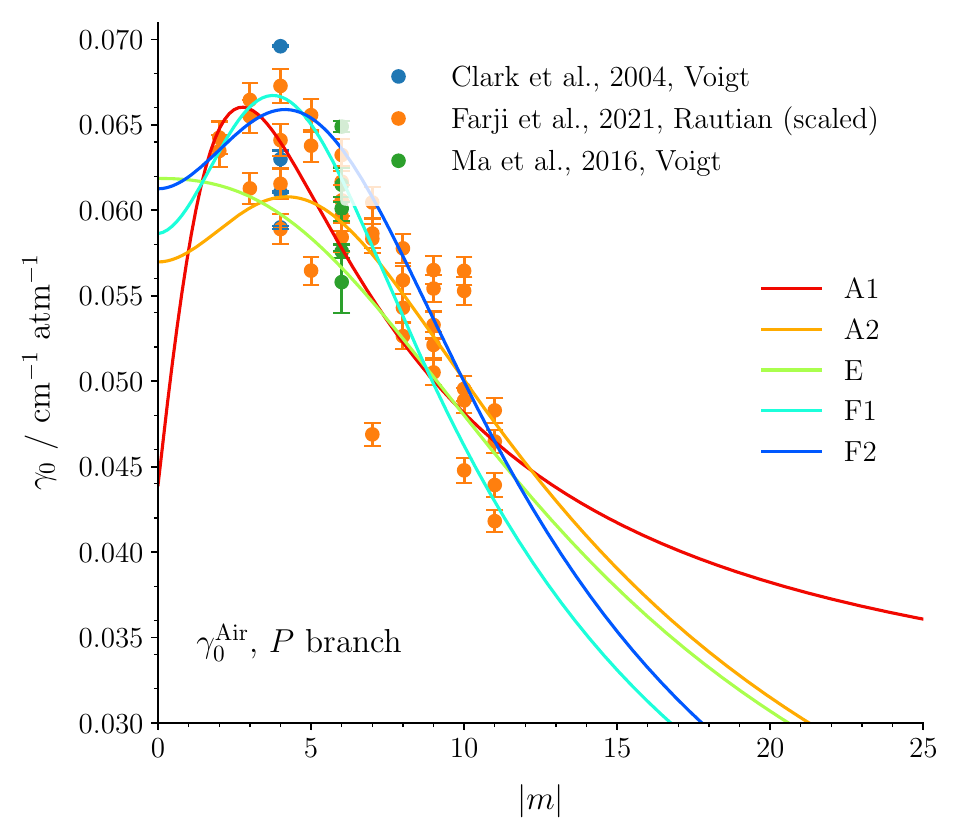}
        \caption{Air-broadening data used for the fit, and resulting Padé approximant for all symmetries}
        \label{fig:pade_fit_allsym}
    \end{figure}
    \begin{table}[htbp]
        \caption{Padé coefficients of each symmetry fit to the parameters found in literature}
        \label{table:pade_parameters}
        \begin{tabular}{lrrrrr}
\toprule
 & \thead[c]{$A1$} & \thead[c]{$A2$} & \thead[c]{$E$} & \thead[c]{$F1$} & \thead[c]{$F2$}\\
\midrule
 $a_0$ & $4.39417 \times 10^{-2}$ & $5.69782 \times 10^{-2}$ & $6.18686 \times 10^{-2}$ & $5.86367 \times 10^{-2}$ & $6.12594 \times 10^{-2}$ \\
\midrule
 $a_1$ & $1.60451 \times 10^{-2}$ & $0.00000$ & $0.00000$ & $2.02531 \times 10^{-4}$ & $0.00000$ \\
\midrule
 $a_2$ & $2.06606 \times 10^{-3}$ & $1.38168 \times 10^{-3}$ & $3.11628 \times 10^{-4}$ & $2.46340 \times 10^{-3}$ & $1.07334 \times 10^{-3}$ \\
\midrule
 $b_1$ & $0.00000$ & $0.00000$ & $0.00000$ & $0.00000$ & $0.00000$ \\
\midrule
 $b_2$ & $7.54018 \times 10^{-2}$ & $1.20675 \times 10^{-2}$ & $6.08344 \times 10^{-3}$ & $1.27335 \times 10^{-2}$ & $4.01121 \times 10^{-3}$ \\
\midrule
 $b_3$ & $7.49201 \times 10^{-16}$ & $1.69122 \times 10^{-3}$ & $3.30447 \times 10^{-4}$ & $4.36874 \times 10^{-3}$ & $1.97354 \times 10^{-3}$ \\
\bottomrule
\end{tabular}

    \end{table}
    The results ({\color{blue}Table \ref{table:pade_parameters}} and Fig. \ref{fig:pade_fit_pr} and \ref{fig:pade_fit_allsym}) yield only a single function per symmetry. However, given the lack of data at higher quanta and ranking indices, these are the only reliable Padé curves that could be calculated. Nevertheless, they should still provide a good consistent set of parameters for spectroscopic observations. The $A_1$ symmetry diverges rapidly from the others. This could simply be a problem with the fit, but again, no assumption at higher quanta should be made, therefore, we decided to keep it.
    The choice of parameters added to HITRAN2024 was based on the data available. Not every transition used the Padé value. Specifically, if high-precision measurements were available, they were taken instead. This rule holds not only for new data, but those present in HITRAN2020 as well, which were not replaced if the uncertainty is lower. With this approach, we are able to conserve the structures observed in the experiments. The air-broadening half-widths of the $\nu _3$ band from HITRAN2020 were replaced using this rule, ensuring a traceable set of data for future updates. Some {\color{blue} previous} values were found to be unrealistically low or high, possibly caused by typographical errors, and were thus replaced as well. Using this approach, the total number of replaced broadening parameters was 30,000.  

    \subsection{\texorpdfstring{\textsuperscript{12}CH\textsubscript{3}D \& \textsuperscript{13}CH\textsubscript{3}D}{12CH3D \& 13CH3D}}

    With its different symmetry, CH\textsubscript{3}D must be treated differently, and benefits from a more rigorous separation based on quantum numbers. The semi-empirical approach developed by Dudaryonok \textit{et al.} \cite{dudaryonok_semiempirical_2018} seems to produce very good results in the $\nu _6$ band. Ignoring the wavenumber dependence, we applied their method for the broadening and temperature-dependence parameters of every new line added in Sec. \ref{sec:addition_shifts}. The wavenumber dependence of the pressure shift was reintroduced, summing the semi-empirical value to the empirical formula in HITRAN2012 \cite{brown_methane_2013} of $-2 \times 10^{-6} \times \Delta\nu$, limited to negative values. A similar approach was used where the extremely low or high values of HITRAN2020 were replaced along with the new lines, which only amount to about 70.

\section{Conclusion}
The parameters of more than 80,000 lines of \textsuperscript{12}CH\textsubscript{4}, \textsuperscript{13}CH\textsubscript{4}, and \textsuperscript{12}CH\textsubscript{3}D have been updated for HITRAN20204 using the new line lists provided. Transitions wavenumbers, intensities, and lower-state energy levels were taken directly from these lists to replace the values from HITRAN2020. Laboratory spectra in the pentad, octad {\color{blue} and tetradecad} regions were used by the HITRAN team to validate and choose between the overlapping data sets, while other regions relied on the validations from the other groups. {\color{blue} We observed a great drop in the residuals of the octad and tetradecad regions spectra particularly thanks to the atm.161 list and the new data from Reed \textit{et al.} \cite{reed_multilaboratory_2025} and Yin \textit{et al.} \cite{yin_2025}. The effects on the pentad region were limited, but still showed improvements, especially at high er energies.} 45,000 lines of the various isotopologues have been added, including some lower-intensity CH\textsubscript{3}D lines that {\color{blue} should have been in the previous update but} were missing, {\color{blue} and increasing the region covered from 12,000 to 14,000 cm$^{-1}$ with experimental data}. In regions that were previously empirically determined, this work allowed us to reassign some of the intensities to the correct isotopologues. Assignments of many lines have been completed between the octad and triacontad regions.

The broadening parameters of these newly added lines were replaced using Padé-approximant functions fit from multiple recent studies for \textsuperscript{12}CH\textsubscript{4} and \textsuperscript{13}CH\textsubscript{4}, unless reliable experimental data were provided, such as those generated at the USTC. The functions were obtained from data sets separated based on their rotational symmetry and ranking index. Currently, only one function per symmetry was used, regardless of the ranking index. The pressure-induced shifts were calculated using the same formula from HITRAN2012 \cite{brown_methane_2013}. Semi-empirical functions from Dudaryonok \textit{et al.} \cite{dudaryonok_semiempirical_2018} were used for \textsuperscript{12}CH\textsubscript{3}D. The pressure shifts were scaled using the HITRAN2012 formula to account for the vibrational dependence, while the broadening parameters were taken as is. The broadening parameters from the $\nu _3$ band, were replaced similarly using the same Padé approximants, however, experimental values were used when available.

In this work, line positions generated from the MARVEL energy levels were utilized in many regions; however, more validations will need to be carried out before updating the line positions in other regions. One of the obvious sources of potential improvement would be the $\nu_4$ band region, where very high-precision experiments were recently carried out \cite{germann_methane_2022} (and included in MARVEL analyses \cite{kefala_empirical_2024}). In the region of the 2$\nu_3$ band, while a major update has been carried out for R and Q branches, the P-branch can still be improved further. In addition, the 3$\nu_3$--2$\nu_3$ hot band can be updated using data derived from recent experiments described in Refs. \cite{foltynowicz_2021,10.1063/5.0223447}.
In future updates, it will also be important to focus on the higher-energy lines, where most of the assignments and energy levels are still missing. New variational predictions \cite{nikitin_pes_2025} of the origins for highly excited methane bands up to 14000 cm$^{-1}$ using redundant coordinates \cite{Nikitin_methane_PES_2016} and the calculation method of \cite{Nikitin_methane_method_2015} could be used to extend the analyses in the future. 
These data are crucial for observations at different temperatures and will allow for better calculations of line broadening, which requires dependence on both $J$ and rotational symmetry (and $K$ in the case of CH\textsubscript{3}D). More work is needed to improve residuals in all spectral regions, especially those relevant to remote observations such as the octad. New work building on this update should allow for better determination and calculations of parameters. Finally, the Padé-approximants method should be expanded to other bands using the new parameters provided in this version. Precise measurements at higher quanta would also improve the accuracy and predictive capability of the model. These data would eliminate the uncertain origin of all the broadening parameters, and perhaps enable the determination of the vibrational dependence.

The relational structure of the HITRAN database \cite{Hill2016} and functionality of HAPI \cite{Kochanov2016} allow one to add non-Voigt line shape parameters. For instance, in this edition, we have added the speed-dependent Voigt parameters and first-order line mixing for air-broadening in the tetradecad region. It will be important to have a consistent set of parameters in all the bands as was done for CO\textsubscript{2} \cite{Hashemi2020}, N\textsubscript{2}O, and CO \cite{Hashemi2021} in HITRAN2020. Of course, it is much harder for methane as there are strong symmetry and vibrational dependencies in comparison with linear molecules. Moreover, it will be important to introduce the full line mixing parameters, where possible. The accuracy of first-order line mixing deteriorates quickly at pressures approaching $1$ atm and above. Dense spectral regions such as $Q$ branches are especially affected and need full line mixing, possibly in combination with speed-dependent effects \cite{ciurylo_speeddependent_2000,bertin_co2_2024}, to be properly measured. Unlike some other parameters, these require complete knowledge of the assignments to be implemented; therefore, more theoretical work will have to be done in preparation. Currently, data from Tran \textit{et al.} is available for CH\textsubscript{4}--air \cite{Tran2006,Tran2010} and other collision partners \cite{tran_model_2006-1,Tran2022}, and will be considered in future updates.

In addition, in order to support the studies of planetary and brown dwarf atmospheres, it is important to provide parameters associated with broadening by ``planetary'' gases, including H$_2$O \cite{Tan2019}, H$_2$, He and CO$_2$ \cite{Tan2022}. Unlike many other gases for which these ``planetary'' broadening parameters are available, only broadening by water vapor is available for methane. In this work, we have used the scaling factors recommended in Ref. \cite{Tan2019} to assign broadening by water vapor values to all new lines of methane. We have collected existing information on broadening by hydrogen, helium, and carbon dioxide (including Refs. \cite{bertin_co2_2024,Tran2022, Varanasi1971,Varanasi1974,Varanasi1990,Fox1989,Gabard2013,es-sebbar_intensities_2014,gharib-nezhad_h2induced_2019,Vispoel2019,Sung2020,Yousefi2021,Hosokawa2025,Clment2025}, but it is still a lot of work to process these data, which often do not agree with each other within stated uncertainties. 

\section*{Author contributions}
\textbf{T. Bertin}: Methodology, Software, Validation, Investigation, Resources, Data Curation, Writing -- Original Draft, Writing -- Review \& Editing, Visualization.
\textbf{I. E. Gordon}: Conceptualization, Methodology, Investigation, Resources, Data Curation, Writing -- Original Draft, Writing -- Review \& Editing, Supervision, Project administration, Funding acquisition.
\textbf{R. J. Hargreaves}: Resources, Data Curation, Writing -- Original Draft, Writing -- Review \& Editing.
\textbf{J. Tennyson}: Resources, Writing -- Original Draft, Writing -- Review \& Editing.
\textbf{S. N. Yurchenko}: Resources, Writing -- Original Draft, Writing -- Review \& Editing.
\textbf{K. Kefala}: Resources.
\textbf{V. Boudon}: Resources, Writing -- Original Draft.
\textbf{C. Richard}: Resources, Writing -- Original Draft.
\textbf{A. V. Nikitin}: Resources, Writing -- Original Draft.
\textbf{V. G. Tyuterev}: Resources, Writing -- Original Draft, Writing -- Review \& Editing.
\textbf{M. Rey}: Resources, Writing -- Original Draft, Writing -- Review \& Editing.
\textbf{M. Birk}: Resources.
\textbf{G. Wagner}: Resources.
\textbf{K. Sung}: Resources, Writing -- Original Draft, Visualization.
\textbf{B. P. Coy}: Resources.
\textbf{W. Broussard}: Resources.
\textbf{G. C. Toon}: Validation, Investigation, Resources, Writing -- Review \& Editing.
\textbf{A. A. Rodina}: Resources.
\textbf{E. Starikova}: Resources, Writing -- Original Draft, Visualization.
\textbf{A. Campargue}: Resources, Writing -- Original Draft, Writing -- Review \& Editing, Visualization.
\textbf{Z. D. Reed}: Resources, Writing -- Original Draft, Writing -- Review \& Editing, Visualization.
\textbf{J. T. Hodges}: Resources, Writing -- Original Draft, Writing -- Review \& Editing.
\textbf{Y. Tan}: Resources, Writing -- Original Draft, Writing -- Review \& Editing, Visualization.
\textbf{N. A. Malarich}: Resources.
\textbf{G. B. Rieker}: Resources, Writing -- Original Draft, Visualization.

\section*{Acknowledgments}
The update of the methane line list in HITRAN is funded through NASA grants 80NSSC23K1596, and 80NSSC24K0080.
Work at UCL was supported by  ERC Advanced Investigator Project 883830; we thank Jingxin Zhang for help generating the HITRAN-style MM line list.

The research at the Jet Propulsion Laboratory, California Institute of Technology, was performed under contract with the National Aeronautics and Space Administration. K. Sung acknowledges JPL Summer Internship Program support for B. P. Coy and A. W. Broussard. Partial support from NASA-ROSES-XRP and from RSF-ANR TEMMEX Russian-French project. Research at the National Institute of Standards and Technology was supported in part by the NIST Greenhouse Gas and Climate Science Program.

\section*{Disclaimer}
Identification of certain commercial equipment, instruments, software, or materials does not imply recommendation or endorsement by the National Institute of Standards and Technology, nor does it imply that the products identified are necessarily the best available for the purpose.

    \bibliographystyle{model3-num-names}
    \biboptions{sort&compress}
    \bibliography{bib_ch4}

\appendix

\section{Abbreviations}\label{append_abbrev}

 The following abbreviations and acronyms have been used throughout the article:

\begin{itemize}
  \item CRDS -- Cavity ring-down spectroscopy
  \item DLR -- Deutsches Zentrum f\"{u}r Luft und Raumfahrt (German Aerospace Center)
  \item FTIR -- Fourier transform infrared
  \item FTS -- Fourier transform spectrometer
  \item GOSAT --  Greenhouse Gases Observing Satellite
  \item HAPI -- HITRAN Application Programming Interface
  \item HITRAN -- High-resolution transmission molecular absorption database
  \item HITEMP -- High-temperature molecular spectroscopic database
  \item HT -- Hartmann--Tran
  \item IAO -- Institute of Atmospheric Optics
  \item ICB -- Laboratoire de l'Universit\'{e} de Bourgogne
  \item JPL -- Jet Propulsion Laboratory
  \item LM -- Line mixing
  \item LMSDVP -- Line-mixing speed-dependent Voigt profile
  \item MARVEL -- Measured Active Rotational-Vibrational Energy Levels
  \item MeCaSDa -- Methane Calculated Spectroscopic Database
  \item NASA -- National Aeronautics and Space Administration
  \item NIST -- National Institute of Standards and Technology
  \item pCqSDHC -- partially-Correlated quadratic-Speed-Dependent Hard-Collision
  \item PES -- Potential energy surface
  \item SDVP -- Speed-dependent Voigt profile
  \item SEOM-IAS -- Scientific Exploitation of Operational Missions — Improved Atmospheric Spectroscopy Databases
  \item SNR -- Signal to noise ratio
  \item TCCON -- Total Carbon Column Observing Network
  \item TheoReTS -- Theoretical Reims-Tomsk Spectral data computed from potential energy and dipole moment surfaces
  \item TROPOMI -- Tropospheric Monitoring Instrument
  \item UCL -- University College London
  \item VAMDC -- Virtual Atomic and Molecular Data Centre
  \item VP -- Voigt profile
\end{itemize}

\end{document}